\documentclass[12pt]{spieman}  
\usepackage{amsmath,amsfonts,amssymb}
\usepackage{graphicx}
\usepackage{setspace}
\usepackage[margin=1in, top=1in, bottom = 1.25in]{geometry}
\usepackage{tocloft}
\usepackage{lipsum}
\usepackage{lineno}
\usepackage{afterpage}
\usepackage{soul} 

\begin{document}


\title{The Dispersion Leverage Coronagraph (DLC): \\ A nulling coronagraph for use on primary objective grating telescopes }

\author[a]{Leaf Swordy}
\author[a]{Heidi Jo Newberg}
\author[a]{Becket Hill}
\author[b]{Richard K. Barry}
\author[a]{Marina Cousins}
\author[a]{Kerrigan Nish}
\author[c]{Frank Ravizza}
\author[a]{Sarah Rickborn}
\affil[a]{Rensselaer Polytechnic Institute, Dept. of Physics, Applied Physics, and Astronomy, 110 8th St, Troy, NY 12180, USA}
\affil[b]{Goddard Space Flight Center, Greenbelt, MD 20771, USA}
\affil[c]{Lawrence Livermore National Laboratory, P.O. Box 808, Livermore, California 94551, USA}
\maketitle

\begin{abstract}
We present the Dispersion Leverage Coronagraph (DLC), a novel variation of the Achromatic Interfero Coronagraph (AIC) that is designed for optical systems featuring large, dispersive primary objective gratings. DLC was originally designed for the Diffractive Interfero Coronagraph Exoplanet Resolver (DICER), a notional 20m class infrared space telescope utilizing the enhanced one-dimensional angular resolution of large diffraction gratings in order to discover and characterize near-Earth exoplanets. Here we develop the theoretical foundation for DLC, and apply it to DICER as an example use case. We derive important properties of the DLC system including focal plane transmission maps, stellar leakage, residual optical path difference tolerance, and pointing error/jitter considerations. Ultimately, we found that DLC effectively nulls an on-axis target across the entire spectrum in the focal plane, allowing for $2D/\lambda$ diffraction-limited imaging. It requires asymmetrical fine-guidance tolerances on pointing error/jitter. We work through a benchmark DICER design, explaining the need for a second disperser to reduce background from Zodiacal light, and showing that it could plausibly find and characterize $\sim 4$ nearby, habitable exoplanets around Sun-like stars in a seven year mission; about 30\% of the habitable exoplanets within 8 pc were found in our simulation. The DLC may be useful for any application requiring extremely high resolution, close-companion spectroscopy.
\end{abstract}


\section{\label{sec:Intro}Introduction}

\subsection{Stellar Interferometry}

Among the earliest attempts to utilize stellar interferometry for high angular resolution observations was the Michelson Stellar Interferometer, which conducted the first measurement of a stellar diameter (Betelgeuse) in 1920\cite{msi}. The first successful aperture-synthesis measurements in the mid-IR using a stellar interferometer were conducted with the Infrared Spatial Interferometer (ISI) in 2003\cite{isi}. Stellar interferometry has since remained a rapidly expanding and evolving field. To date, the smallest inner working angle measurements of circumstellar material have been made using beam combiner instruments such as CHARA/FLUOR\cite{chara}, KECK/KIN\cite{kin}, LBTI/NOMIC\cite{lbti}, VLTI/PIONIER\cite{ertel2018}, and the planned VLTI/NOTT\cite{NOTT}. Space based interferometric nulling concepts for resolving close-companions from bright host-stars were first  formulated by R. Bracewell in 1978\cite{BRACEWELL1979136}. The original Bracewell design combines light from two or more apertures and, after applying a $\pi$-phase shift to one half of the optical input, achieves an interferometric null for on-axis light sources. Following the initial Bracewell configuration, many other nulling architectures have been proposed utilizing achromatic pupil/phase inversion\cite{RABBIA2007385}, pupil apodization\cite{apod}, and a variety of phase-masking techniques\cite{fqpm,Mawet_2009}. 

Many current and planned science missions utilize nulling interferometry in order to study exozodiacal dust, protoplanetary disks, and near-Earth exoplanets. Ground-based implementations of nulling interferometry are often contextualized as precursor missions to future space observatories, capable of directly resolving nearby Earth-like exoplanets in the Mid infrared (MIR). Missions attempting to capitalize on this science objective include the canceled Terrestrial Planet Finder\cite{TPF} (TPF) and Darwin\cite{LUND2001137} missions, along with the more contemporary Large Interferometer For Exoplanets\cite{refId0,refId2} (LIFE), which reexamines exoplanet resolving with space interferometry.

In recent years, many notional telescope architectures\cite{Ditto03} utilizing Primary Objective Gratings (POGs) for optical leverage have been formulated. Such designs aim to utilize diffraction gratings, as well as other holographic primaries, for a host of potential enhancements over conventional mirrors. These enhancements include improved angular resolution, decreased aerial mass, and increased field of view. They are discussed in detail in Swordy, Newberg, \& Ditto (2023)\cite{tel}.

We describe here the Dispersion Leverage Coronagraph (DLC), a novel coronagraph architecture which can be used on a POG telescope for bright source suppression. An earlier version of this work was previously published in the proceedings of SPIE in Swordy et al. (2023)\cite{SPIEDLC}. The work presented here includes updates and corrections to this previous publication, along with an example application the Diffractive Interfero Coronagraph Exoplanet Resolver (DICER)\cite{dicer2}. 

The fundamental mode of operation of DLC is very similar to Achromatic Interfero Coronagraphy\cite{RABBIA2007385} (AIC), which nulls a central source by passing half of a beam through focus. However, the dispersion and coordinate asymmetry introduced by the POG leads to a very unique optical system in need of theoretical characterization. This work gives an overview of the DLC optical system, and aims to establish the theoretical foundations for DLC, with an emphasis on standard coronagraph metrics such as the transmission map, stellar leakage, tolerance to pointing jitter, and residual Optical Path Difference (OPD) leakage. Detailed numerical discussions of close-companion detection efficiency are highly application specific, but we outline the parameters for one application that has been developed, DICER, as an example.

\subsection{\label{sec:dicer} POG Telescopy}

\begin{figure}
\begin{center}
\begin{tabular}{c}
\includegraphics[height=6.5cm]{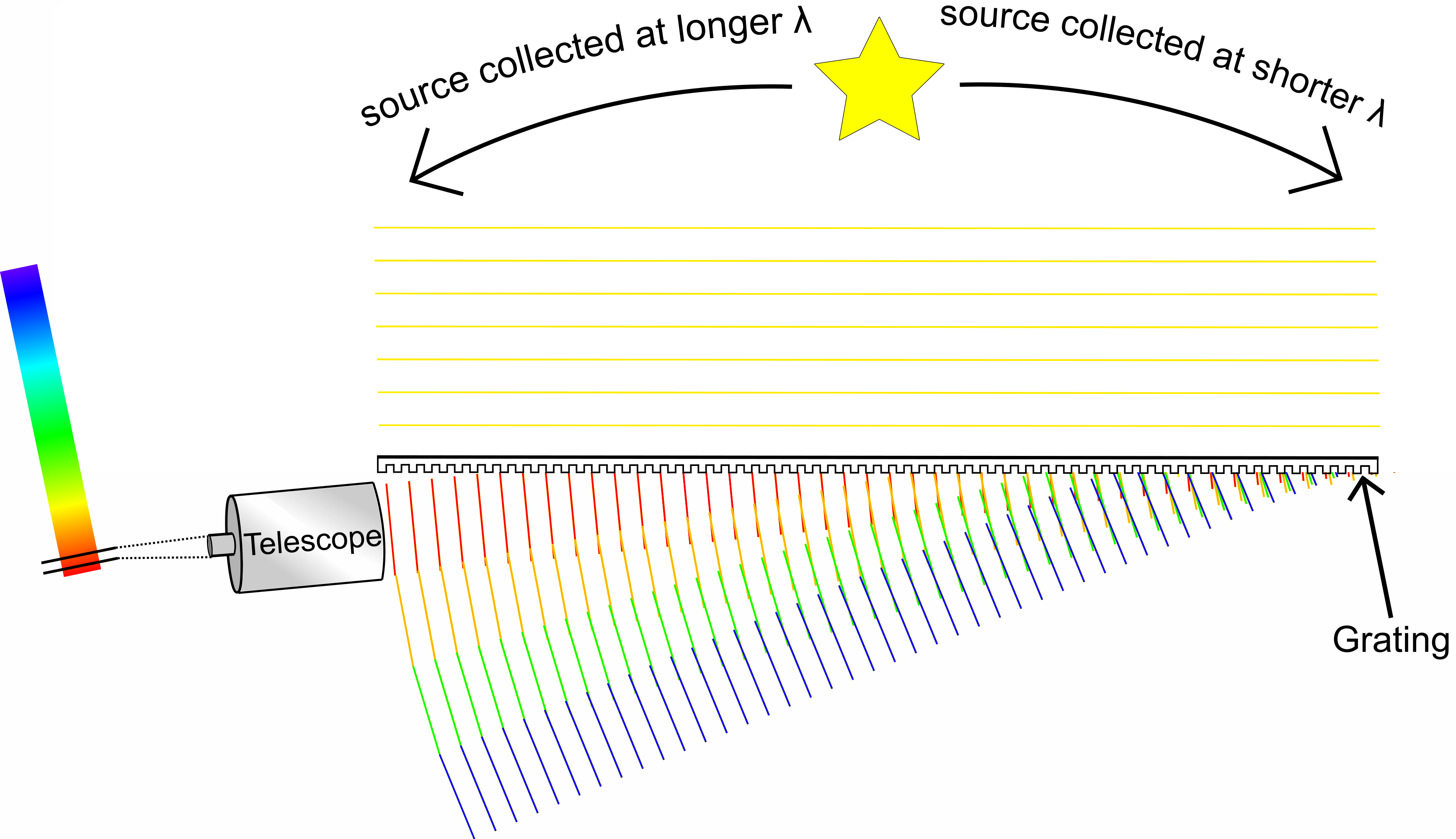}
\\
\end{tabular}
\end{center}
\caption 
{ \label{fig:trace}
Placing a telescope at an extreme angle to a diffraction grating allows the telescope to see the entire length of the grating while collecting a small slice of the overall grating output spectrum. A source is depicted above the grating, with wavefronts from the source arriving parallel to the grating surface. The curved arrows indicate that if the source position is shifted in the sky, the thin slice of collected spectrum shifts in wavelength.
}
\end{figure} 

A POG telescope is, at the most basic level, composed of a focusing element placed at the output of a large diffraction grating in order to capture a component of the output spectrum. If the telescope is placed at an extreme angle to the grating (Figure \ref{fig:trace}) a very large grating area can be observed by a comparatively small focusing element.

A key characteristic of POG telescopes is the correlation between received wavelength and the angular position of objects on the sky, as shown in Figure \ref{fig:trace}. The angular resolution of a POG telescope is directly related to the spectral resolution of the illuminated grating area.\cite{tel} Specifically, if $\theta_{in}$ and $\theta_{out}$ represent the input and output angles to the grating respectively, we can derive the following anamorphic magnification relation by integrating the grating equation:

\begin{equation}
\frac{d\theta_{out}}{d\theta_{in}} = \frac{\textrm{cos}(\theta_{in})}{\textrm{cos}(\theta_{out})},
\end{equation}

\noindent This equation is written in the notation of Reference 18 (where a more complete discussion of these topics can be found), which will differ from the notation used later in this text.

Because the grating shape must be much longer in the dispersion direction for the telescope to capture light at near grazing exodus, this spectral/angular resolution is only enhanced in one dimension compared to the telescope resolution without the grating, and scales linearly with the grating length in a similar manner to the Rayleigh criterion for conventional telescopes.

\subsection{DICER}

\begin{figure}
\begin{center}
\begin{tabular}{c}
\includegraphics[width=0.9\linewidth]{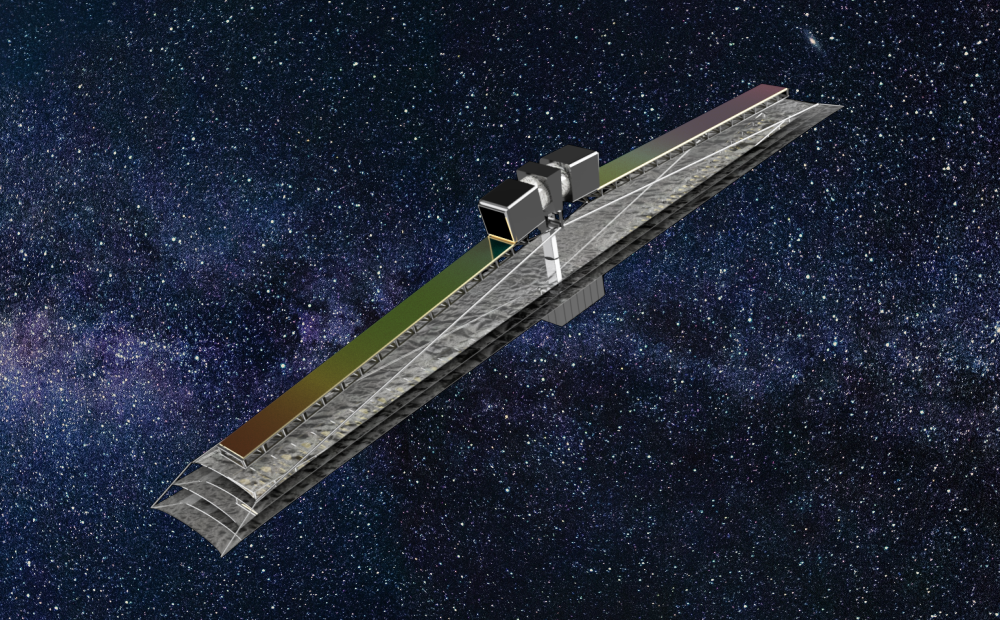}
\\
\end{tabular}
\end{center}
\caption 
{ \label{fig:DICER}
Design concept for an infrared observatory at the Earth-Sun L2 point that collects light with two large POGs (shown in rainbow colors), which are observed by secondary optics similar to a conventional infrared telescope mirror. Light from the two telescopes is combined in a DLC coronagraph that nulls the exoplanet’s host star, enabling detection of small segments of an exoplanet’s emission spectrum.
}
\end{figure}

In this paper, we will use the example of the Diffractive Interfero Coronagraph Exoplanet Resolver
\cite{10.1117/12.2629487} (DICER) to illustrate the use of DLC. DICER (Figure \ref{fig:DICER}) utilizes low aerial mass diffraction gratings to achieve the angular resolution of a 20m aperture. Operating near the peak of the Earth's infrared emission spectrum in the vicinity of 10$\mu$m (where the Earth/Sun contrast ratio is most favorable), DICER seeks to directly resolve warm terrestrial exoplanets in the habitable zone. The one-dimensional angular resolution of DICER allows for exoplanet discovery/detection including approximations of orbit inclination and semi-major axis. The dispersive optics of DICER also allow for detection of biomarkers in spectra of exoplanet atmospheres; the strong ozone absorption line just shy of 10$\mu$m is a primary science driver.

We show here that the DICER design, while complex, could actually be used to observe exoplanets. DICER takes advantage of the enhanced one-dimensional angular resolution of the POG to resolve Earth-like exoplanets from the Sun-like stars they are orbiting out to a distance of 10pc. The notional DICER design features two linearly-aligned diffraction gratings that feed into secondary telescopes, which then pass the light into the DLC interferometer (Figures \ref{fig:aicfig}a \& \ref{fig:pog}). By observing in the vicinity of 10$\mu$m (near the peak of Earth's blackbody emission), DICER can directly resolve a small slice of the exoplanet emission spectrum around the prominent ozone absorption line just shy of 10$\mu$m.\cite{10.1117/12.2629487}

In this paper, we will present a benchmark design for DICER, that plausibly illustrates that DICER could detect exoplanets. However, we will also show that the low luminosity of habitable exoplanets and the high backgrounds of zodiacal light make the engineering requirements unusually challenging. Any Dittoscope design faces engineering challenges that are directly related to the primary objective gratings, which in this case are 10m in length. The grating must have a tightly constrained differential line density, which requires manufacturing precision and thermal stability sufficient to maintain a regular line spacing across the grating (or grating segment, if the 10m grating is composed of multiple, aligned gratings). The need to reduce zodiacal light background necessitates a very complex spectrograph in the focal plane.

However, exoplanets are not the only science use case for which DLC could potentially be used. Other applications that are under exploration include identification of binary stars and stellar mass loss on the giant branch. In the discussion, we present applications that are less demanding than the habitable exoplanet illustration. Less demanding applications might require smaller gratings, brighter sources, or optical wavelength bands. While we do discuss some specific DLC science cases, this work mainly aims to establish the theoretical foundations of DLC in all generality, regardless of application.

\section{\label{sec:config}DLC Configuration}

\begin{figure}
\begin{center}
\begin{tabular}{c}
\includegraphics[height=6cm]{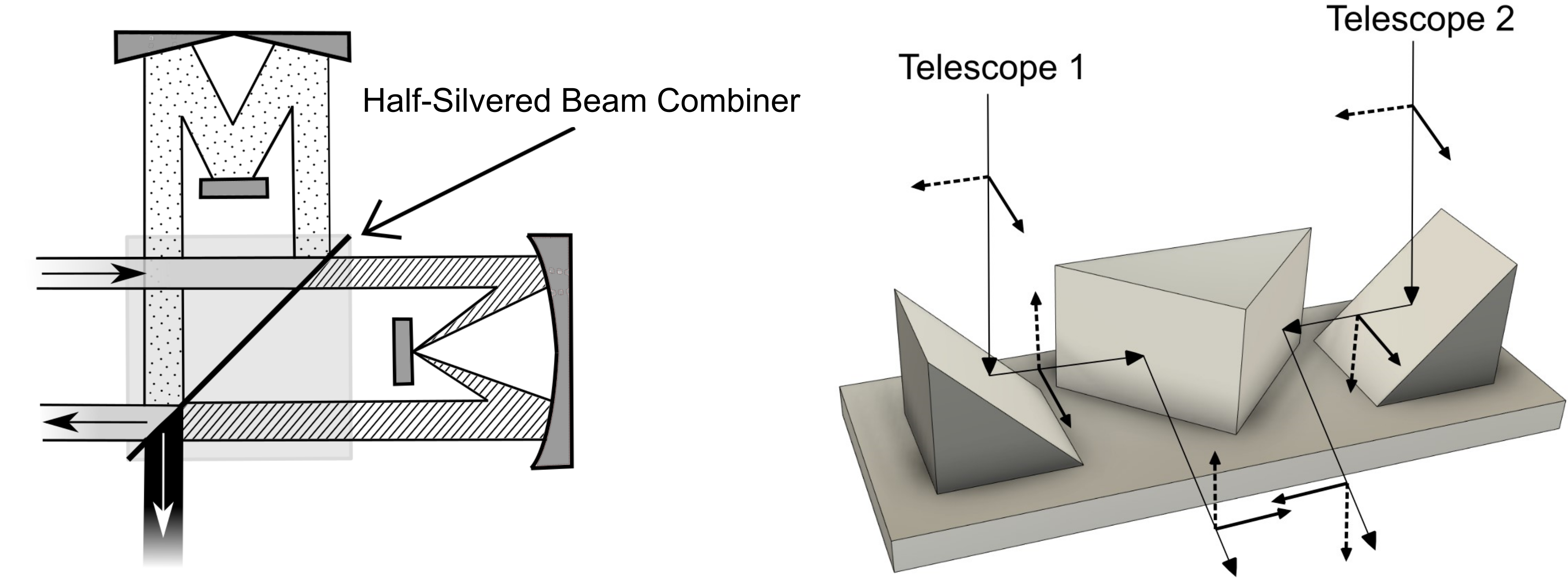}
\end{tabular}
\end{center}
\caption 
{ \label{fig:aicfig}
(a) Optical path diagram of the AIC interferometer in the case of splitting the output from a single telescope. Light entering from the center-left is split between two paths. The right path undergoes a $\pi$ phase shift and pupil inversion after crossing focus, while the upper path is unchanged. On recombination, the bottom output destructively interferes. On an AIC equipped telescope, the bottom output is relayed to the focal plane. As a result, all centro-symmetric incident light is rejected by the interferometer. (b) Schematic diagram of the periscope nulling scheme used by DLC. Light from two telescopes (telescopes 1\&2, gathering the diffracted light from gratings 1\&2) is passed through the periscope monolith pictured above. Field vectors of the input beams are represented as solid/dotted arrows. The two wings of the periscope exhibit a left/right hand asymmetry, resulting in an achromatic $\pi$ phase shift and pupil inversion between the two beams.
}
\end{figure} 

AIC combines the input from two separate telescopes (or alternatively a single telescope with split output), after passing through an optical system resembling a Michelson interferometer. A simple diagram of the AIC optical train is shown in Figure \ref{fig:aicfig}a. In the AIC interferometer, host-star light is nulled by application of a centro-symmetric pupil inversion/rotation and $\pi$ phase shift produced by allowing the signal from one wing of the interferometer to cross focus prior to recombination (right path in Figure \ref{fig:aicfig}a). The end result is a focal plane containing a central null corresponding to input sources with zero angle of incidence relative to the telescope optical axis. Off-axis sources are preserved because the phase-shifted companion PSFs do not land at the same location in the focal plane, instead appearing as a double-image of the companion located on either side of the nulled host-star position.

Experimental evaluations of mid-infrared coronagraph nulling architectures\cite{Gappinger:09} have highlighted the difficulty of achieving deep starlight suppression using the through-focus method of phase/pupil inversion. Such studies suggest a generally superior technique, the nulling periscope (Fig. \ref{fig:aicfig}b), which can be used to achieve the same achromatic phase shift and pupil inversion without the need for powered optics in the interferometer. The periscope nulling scheme utilizes an antisymmetric arrangement of flat mirrors to achieve an achromatic electric field flip. This field flip results in a $\pi$ phase shift pupil inversion that is virtually identical to that achieved using the through-focus method, but is generally much easier to align and exhibits fewer phase defects arising from powered interferometer components.

\begin{figure}
\begin{center}
\begin{tabular}{c}
\includegraphics[height=5.6cm]{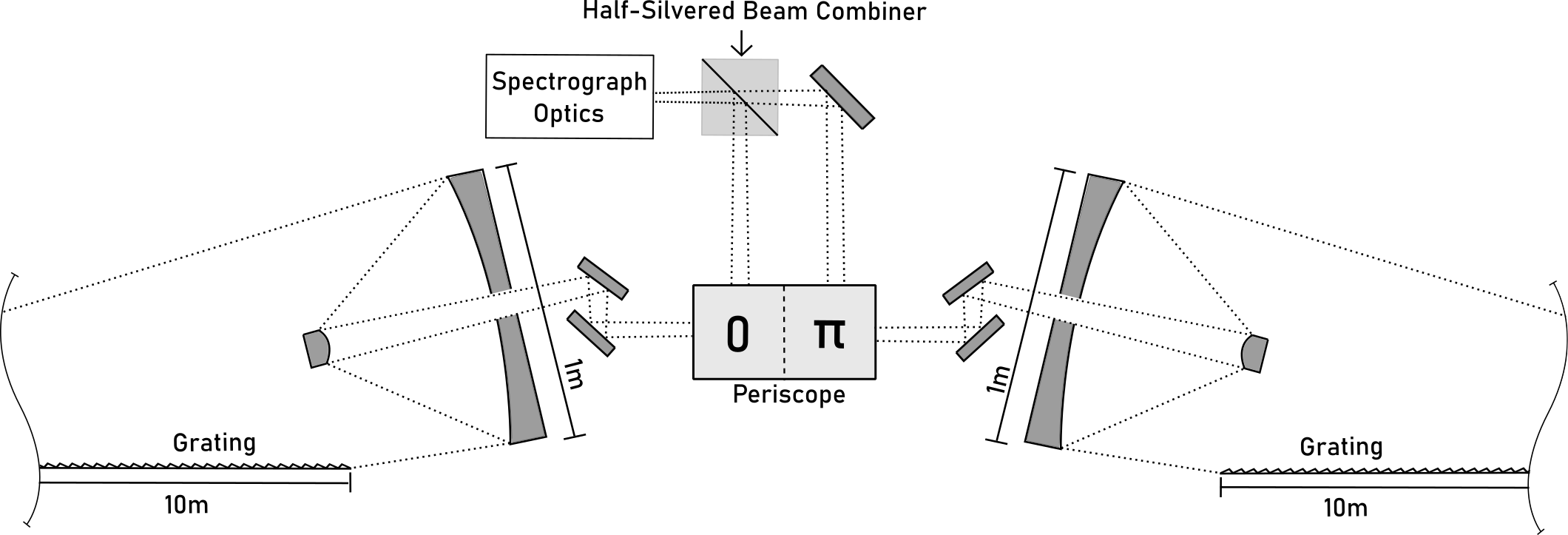}
\end{tabular}
\end{center}
\caption 
{ \label{fig:pog}
Optical path diagram of the DLC optical train. Light collected from two gratings (either reflection or transmission) is directed into a central interferometer utilizing the field-flipping periscope. The light from each telescope is ultimately recombined and imaged following an achromatic $\pi$ phase shift and pupil inversion introduced by the periscope optics.
}
\end{figure} 

\begin{figure}[h!]
\begin{center}
\includegraphics[width=1.0\linewidth]{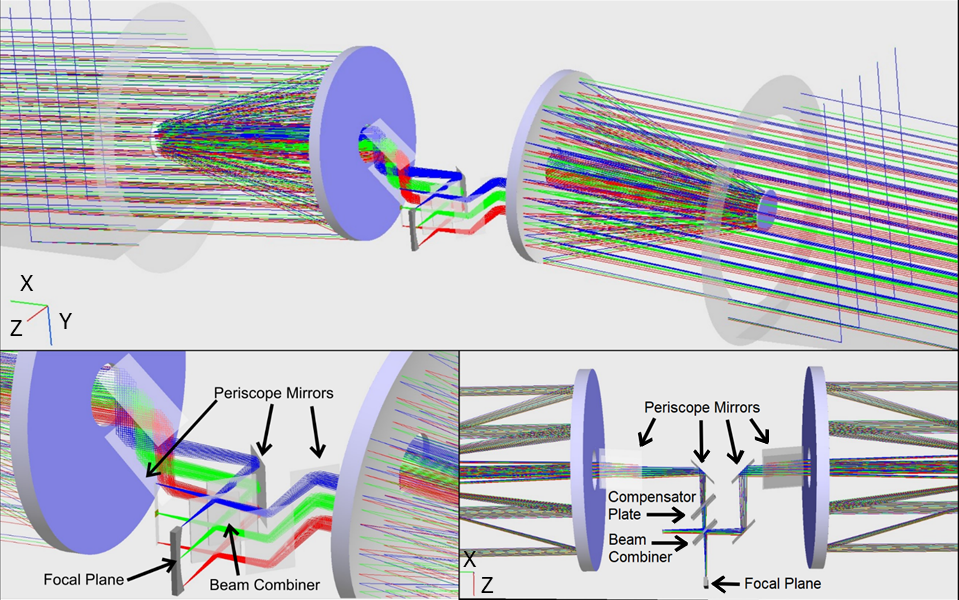}
\end{center}
\caption{Ray trace of DLC. The top panel shows light of three different wavelengths being diffracted by two flat gratings, and then focused by beam-reducing optics with a Ritchey-Chr\'etien telescope design before going through periscope optics, a beam combiner, and then being focused to a curved focal plane. The beam reducer is modeled with a 15 meter focal length, 2.5 meter back focal length, and a primary-secondary distance of 1.2 meters. The colored traces represent light with wavelengths that are slightly longer (red) and shorter (blue) than the central 10$\mu$m wavelength depicted in green. Light from the left (+X) beam reducer is reflected down (+Y), then towards the focal plane (+Z) and subsequently passes through a compensator plate and a half-silvered beam combiner before reaching the focal plane. Light from the right (-X) beam reducer is reflected down (+Y), and then is reflected off two periscope mirrors, passes through the beam combiner, is reflected back for a second pass through the combiner before impinging on the focal plane. Light from the +X beam reducer only passes through the beam combiner plate one time, which is why it must pass through a compensator plate to have the same path length to the focal plane as the light from the right. The lower left panel shows an enlarged view of the periscope mirrors and DLC focal plane (with the same orientation as the top panel). The lower right shows a rotated view of the periscope that makes it easier to identify the DLC optical components. In practice, the interferometer portion of the optical train would require active cooling and electromechanical path length compensation rather than a compensator plate in one beam to maintain the accuracy of our theoretical calculations.}\label{fig:DLCraytrace}

\end{figure}

The DLC concept takes the output from two POGs at grazing exodus and feeds them into a nulling periscope interferometer (Figure \ref{fig:pog}). Figure \ref{fig:DLCraytrace} shows a ray trace using Photon Engineering's FRED Optical Engineering Software of three different wavelengths of light that have been diffracted by the grating and are then focused by a secondary optic that is of Ritchey Chr\'{e}tien design. For ease of coding in FRED, the secondary telescopes are cylindrically symmetric, in contrast to the benchmark design, which has square secondary telescopes as pictured in Figure~\ref{fig:DICER}. 

\begin{figure}[h]
\begin{center}
\begin{tabular}{c}
\includegraphics[width=0.7\linewidth]{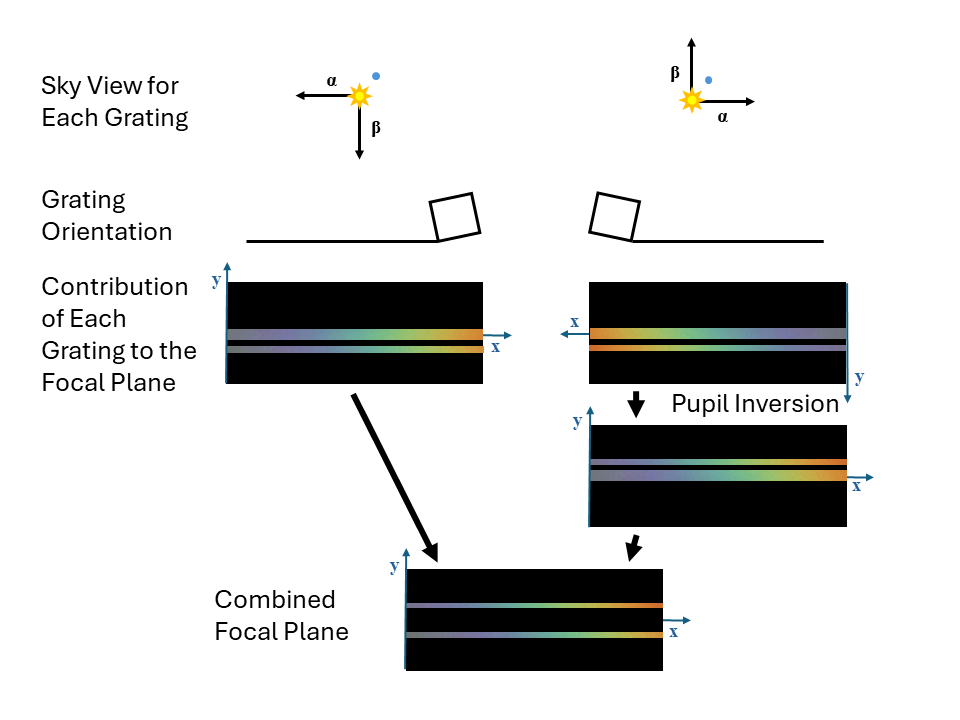}
\end{tabular}
\end{center}
\caption 
{ \label{fig:coords}
Schematic diagram demonstrating how the sky image as viewed by the two telescopes with primary objective gratings is conveyed to the final combined focal plane.  Because the two grating (horizontal black line) plus secondary telescope systems (attached tilted rectangle) are oriented $180^\circ$ to each other, if the gratings are pointed exactly at the host star then the angle of incidence of the planet has the opposite sign for each system. The focal planes of these two systems would therefore be flipped from each other, as shown in ``Contribution of Each Grating to the Focal Plane," and would capture different wavelength ranges of the planet. We pupil invert light from one of the systems before combining the beams in the focal plane, so that the host star light is nulled. The exoplanet spectra are shifted in wavelength in the focal plane so that even if the line from the star to the exoplanet is exactly aligned with the long axis of the grating so that the exoplanet spectra from the two sides lie on top of one another, the exoplanet light will not null. Note that since the telescopes are pointed in opposite directions, the pupil inversion serves to line up (in terms of $\lambda$) the spectra in the focal plane. This contrasts with AIC, in which the focal planes of the two telescopes are only aligned in the center (zero angular offset in either direction). Also note that the planet/companion spectrum is shifted either up or down in $\lambda$ depending on which side of the grating normal the object incident angle falls. Although the spectrum is shown in the figure as covering the whole optical wavelength range, DICER only captures a 14.7 nm bandwidth near 10$\mu$m. The pictured $(\alpha, \beta)$ sky coordinates and $(x, y)$ focal plane coordinates are as defined in Figure\ref{fig:diff}b (not to be confused with the $(X, Y)$ coordinates of the FRED simulation).} 
\end{figure}

\begin{figure}
\begin{center}
\begin{tabular}{c}
\includegraphics[width=1\linewidth]{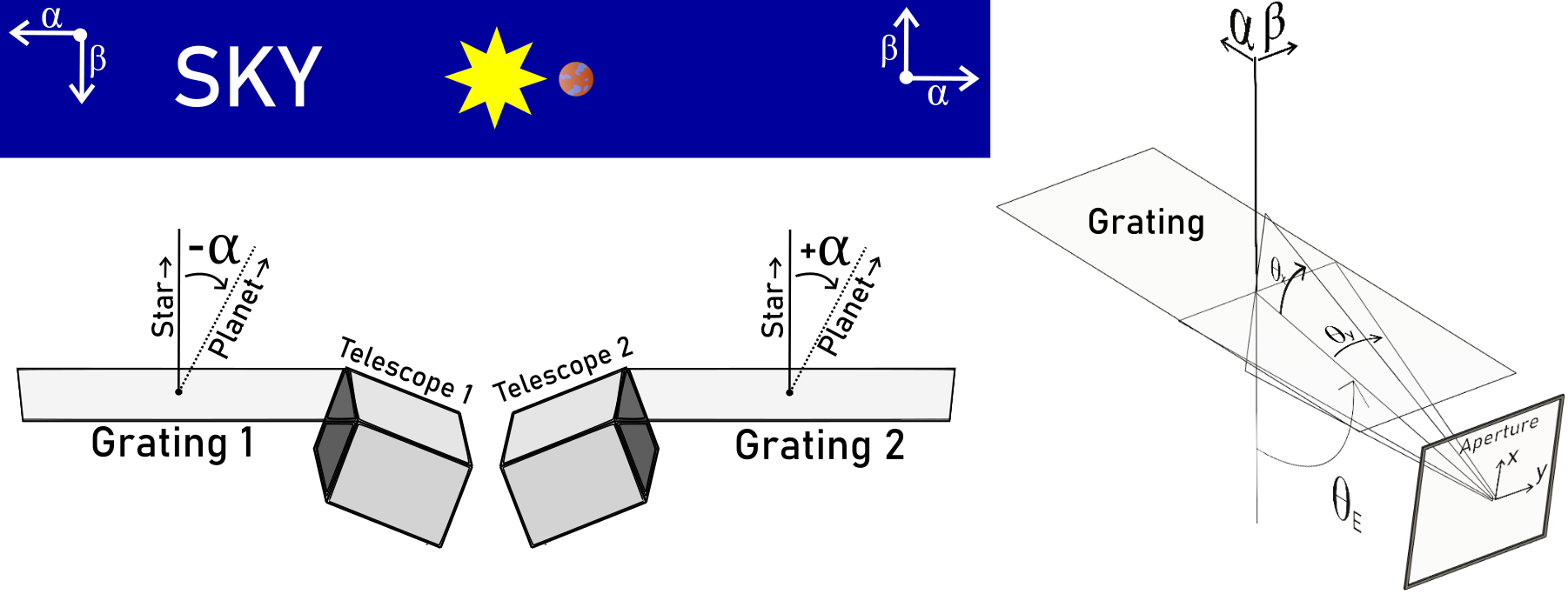}
\end{tabular}
\end{center}
\caption 
{ \label{fig:diff}
(a) Schematic diagram demonstrating that, if the gratings are aligned along the angular separation of the host star and close companion, the angle between the companion input wave and the grating normal ($\alpha$) has either positive or negative sign in the coordinate system of the secondary telescope. The sign difference in $\alpha$ as observed by the two telescopes is pivotal in preserving (not nulling) the close-companion light. This is elaborated upon in Section \ref{sect:sims}. (b) Coordinate system for grating incident angles $\alpha$ $\beta$, secondary telescope incident angles $\theta_x$ $\theta_y$, and the optical axis angle $\theta_E$. This coordinate system is used in the mathematical derivations of Section \ref{sect:math}.
}
\end{figure}

Although DLC utilizes an interferometric system that closely resembles AIC, the system differs from conventional AIC in several important ways. First and foremost, instead of a single host-star PSF and companion doublet, the focal plane contains a single host-star spectrum and two shifted companion spectra (Figure \ref{fig:coords}). The host-star spectrum is composed of the overlapping spectra from the two secondary telescopes, and nulls across the entire focal plane. Nulling across the entire focal plane is possible because the secondary telescopes are pointed in opposite directions while still observing the same object (Figure \ref{fig:diff}). In DLC, the pupil inversion/rotation step of the interferometer serves to `line-up' the host star spectra in the focal plane, allowing the spectrum to be nulled along the entire wavelength-dependent focal plane coordinate ($x$).

 If the gratings are aligned along the line of separation between the host star and close companion, the shifted exoplanet spectra also overlap across the entire focal plane. However, the single-wavelength PSFs are unable to effectively interfere due the minute wavelength shift incurred from differing incidence angle to the grating relative to the secondary telescopes (Figure \ref{fig:diff}). Although each single-wavelength PSF from one secondary telescope will completely overlap with a phase-shifted PSF from the other secondary telescope, the overlapping PSFs will exhibit a beat frequency arising from the wavelength shift after dispersion. Over any appreciably long integration time, the inference beat-frequency will average out to a standard signal. This represents a key departure of DLC from conventional AIC. Close-companion light is preserved due to spectral (as opposed to geometric) asymmetry in the focal plane. In this way, the spectral resolvance of the grating is the determining factor in whether or not close-companion light is received in the focal plane.

\section{\label{sect:math}Intensity Distributions from focal plane complex amplitudes}

This section lays the groundwork for determining all of the useful coronagraphic metrics for DLC (the subject of Sections \ref{sec:tran}-\ref{sect:perf}). First we derive the intensity distribution of a single aperture, with a monochromatic plane wave as input (taken as a single component of the POG output spectrum). The single-aperture intensity function will then be used to derive the overall intensity distribution of the interferometric system, comprised of the two secondary telescopes/gratings and the DLC optical train.

Derivation of the intensity function is much simpler under the assumption of a square aperture, as the integrals will be separable in the focal plane/aperture coordinates $x$ and $y$. This is especially important since unlike most applications featuring circular optics, the optical system exhibits bi-lateral (as opposed to radial) symmetry. The choice of a square aperture may also be natural, since a square aperture is needed to collect the exodus light from a rectangular grating. The coordinate system for all subsequent calculations is taken from the right panel of Figure \ref{fig:diff}.

\subsection{Single Aperture Intensity Distribution}
In the Fraunhofer approximation, incident plane waves arrive at the focal plane after traversing the focal length $f$ (not to be confused with the function $f(x,y)$ defined below) with incident angles $\theta_x \approx \frac{x}{f}$ and $\theta_y \approx \frac{y}{f}$, with corresponding spatial frequencies $v_x= \frac{x}{\lambda f}$, $v_y= \frac{y}{\lambda f}$ in the aperture plane. We model incident light from the host star or companion with complex amplitudes $U(x,y,z) = \sqrt{I_0}e^{-i\bf{k}\cdot \bf{r}} = \sqrt{I_0}e^{-i(k_x x + k_y y + k_z z)}$. From the view of Fourier optics, we define the input (aperture) plane as the incident wave complex amplitude evaluated in the plane $U(x,y,0) = f(x,y) = \sqrt{I_0}e^{-i2\pi(v_x x + v_y y)}$, with spatial frequencies $v_x = k_x/2\pi$, $v_y = k_y/2\pi$.

In the case of Fraunhofer diffraction from a square aperture of side length $D$ (subsequently focused by a lens/mirror of focal length $f$), we may calculate the complex amplitude $g(x,y,0)$ at the output (focal) plane as:

\begin{equation}
g(x,y) \approx  h_0 P(v_x,v_y) =  h_0 P\left( \frac{x}{\lambda f},\frac{y}{\lambda f}\right) \; ,
\end{equation} 

\noindent where $P(v_x,v_y) $ is the Fourier transform of the aperture $a(x,y)$ times the complex amplitude $f(x,y)$ at the input plane $\left( p(x,y) = f(x,y)a(x,y)\right)$, and $h_0 = \frac{i}{\lambda f}e^{-ikf}$ is a constant factor arising from the impulse-response function of free space. Therefore,

\begin{equation}
  h_0 P\left( \frac{x}{\lambda f},\frac{y}{\lambda f}\right)=  h_0  \int_{-\infty}^{\infty} \int_{-\infty}^{\infty} a(x',y') f(x',y')\textrm{exp}\left(i 2 \pi \frac{xx' + yy'}{\lambda f}\right) dx'dy'.
\end{equation} 

The aperture function $a(x,y)$ is simply 1 for values $x$, $y$ in the range $-\frac{D}{2}$ to $\frac{D}{2}$, and zero elsewhere. With the use of this aperture function, and after plugging in the input complex amplitude $f(x,y)$, the above integral is reduced to:

\begin{equation*}
g(x,y) \approx  \sqrt{I_0} h_0  \int_{-D/2}^{D/2}\int_{-D/2}^{D/2} \textrm{exp}\left(-i2\pi( v_{0x}x' + v_{0y}y' )\right) \textrm{exp}\left(i 2 \pi \frac{xx' + yy'}{\lambda f}\right) dx'dy'
\end{equation*}

\begin{equation}
= \sqrt{I_0} h_0  \int_{-D/2}^{D/2}\int_{-D/2}^{D/2}  \textrm{exp} \left( \left( \frac{i 2 \pi x}{\lambda f} -i2\pi v_{0x} \right) x' \right)   \textrm{exp} \left( \left( \frac{i 2 \pi y}{\lambda f} -i2\pi v_{0y} \right) y' \right)  dx'dy'
\end{equation} 

\begin{equation*}
= \sqrt{I_0} h_0 \; \frac{\textrm{exp} \left( \left( \frac{i 2 \pi x}{\lambda f} -i2\pi v_{0x} \right) x' \right)}{\left( \frac{i 2 \pi x}{\lambda f} -i2\pi v_{0x} \right)} \Biggr|_{x'=-D/2}^{x'=D/2} * \frac{\textrm{exp} \left( \left( \frac{i 2 \pi y}{\lambda f} -i2\pi v_{0y} \right) y' \right)}{\left( \frac{i 2 \pi y}{\lambda f} -i2\pi v_{0y} \right)}\Biggr|_{y'=-D/2}^{y'=D/2},
\end{equation*} 

\noindent where $v_{0x}$ and $v_{0y}$ specify the incidence angle of the wave crossing the aperture at the input plane. The above expression reduces to the two-dimensional $\textrm{sinc}$ function, a typical PSF for rectangular apertures:

\begin{equation}
\label{eqn:blah}
g(x,y) = \sqrt{I_0} h_0 D^2 \; \textrm{sinc}\left( \frac{Dx}{\lambda f} - \frac{Dx_0}{\lambda f}  \right)      \textrm{sinc}\left(  \frac{Dy}{\lambda f} - \frac{Dy_0}{\lambda f}  \right),
\end{equation}

\noindent where sinc$(x)$ is the normalized sinc function $\frac{\textrm{sin}(\pi x)}{\pi x}$. The intensity in the focal plane is given by:

\begin{equation}
I(x,y) = |g(x,y)|^2 \; .
\end{equation}

\subsection{Interference of Two Apertures}

Due to the nature of DLC, there will be two (possibly) overlapping PSFs incident on the focal plane, each originating from one of the two telescopes in the arrangement. The incident waves from each telescope $U_0(x,y)$, $U_1(x,y)$ will exhibit spatial frequencies $(v_{0x}$ and $v_{0y})$ and $(v_{1x}$ and $v_{1y})$, with input angles of $\theta_{0,x} \approx \lambda v_{0,x}$ and $\theta_{1,x} \approx \lambda v_{1,x}$ in the plane of dispersion, and $\theta_{0,y} \approx \lambda v_{0,y}$ $\theta_{1,y} \approx \lambda v_{1,y}$ in the oblique (non-dispersing) direction. Furthermore, the output complex amplitude $U_1(x,y)$ will incur a $\pi$ phase shift and pupil inversion prior to incidence on the focal plane. Since the telescopes are pointed in opposite directions, yet still observing the same object (see Figures  \ref{fig:diff} \& \ref{fig:coords}), the pupil inversion serves to line up the spectra in the $x$-direction, while angular offsets in the $y$-direction are reflected across the $x$-axis.

To determine the intensity $I_{tot}(x,y)$ of the interference pattern, we must subtract the output complex amplitudes (the phase factor $e^{i\pi}$ picked up by $U_1$ is simply a factor of -1, so the intensities are subtracted rather than added):

\begin{equation}
I_{tot}(x,y) = |g_0(x,y) - g_1(x,y)|^2 = |g_0(x,y)|^2 + |g_1(x,y)|^2  - g_0(x,y)^*g_1(x,y) - g_0(x,y)g_1(x,y)^*.
\end{equation}

It was seen in Eqn. \ref{eqn:blah} that apart from the factor $h_0$, $g(x,y)$ is a real function. It is also easy to see that the phase factor $h_0$ will result in a constant factor of $(\lambda f)^{-2}$ for every term. $g_0(x,y)$ and $g_1(x,y)$ possess the same functional form, but are centered at different locations in the focal plane, shifted from one another by focal plane displacements $\delta_{0x} = \lambda f v_{0x}$, $\delta_{1x} = \lambda f v_{1x}$, $\delta_{0y} = \lambda f v_{0y}$, and $\delta_{1y} =-\lambda f v_{1y}$. Since the y direction is merely a reflection about $x=0$, $\delta_{1y} = -\delta_{0y}$. All of this taken together results in an intensity distribution given by:

\begin{equation}
\label{eqn:intensdist}
I_{tot}(x,y) =   I(x-\delta_{0x},y-\delta_{0y}) + I(x-\delta_{1x},y+\delta_{0y})  - 2g(x-\delta_{0x},y-\delta_{0y})g(x-\delta_{1x},y+\delta_{0y}).
\end{equation}

The third term in the above expression describes the degree of overlap between the complex amplitudes $g_0(x,y)$ and $g_1(x,y)$. It is clear that if the PSF positional shift in the y direction ($\delta_{0y}$) is equal to zero, and the dispersion direction shifts, $\delta_{0x}$ and $\delta_{1x}$, are equal, the intensity will be identically zero. This describes the situation in which the two PSFs are located at the same position in the focal plane and overlap completely; the relative $\pi$ phase shift results in complete destructive interference. If the gratings are pointed directly at the host star, then its light will completely destructively interfere (not considering, for the moment, the issue of stellar leakage).

For the companion, $\delta_{0y}$ will not equal zero unless the grating direction is aligned with the angular separation of the star and companion. If the diffraction-limited resolution of the secondary telescope is small compared to the diffraction limit of the grating, $\delta_{0y}$ will be very small for all orientations. The terms $\delta_{0x}$ and $\delta_{1x}$ are determined by the output angles of the grating, and will only be equal for plane waves with equal incidence angle on the two primary objective gratings.

\section{\label{sec:tran}Single-Wavelength Transmission Map of DLC}

In order to determine the total transmission of an input wave at the focal plane, the total intensity $I_{tot}(x,y)$ of Equation \ref{eqn:intensdist} must be integrated over the entire focal plane. Luckily, the $x$- and $y$-dependent terms of the functions $g(x,y)$ that combine to construct $I_{tot}(x,y)$ are separable in the focal plane coordinates $x$ and $y$. The total transmission of DLC is found by performing integrals of the form found in Appendix \ref{sect:ident} over the entire focal plane:

\begin{equation}
\label{eqn:wholef}
\int_{-\infty}^\infty \textrm{sinc}(bx+c) \textrm{sinc}(bx+a)dx = \frac{\textrm{sin}(\pi(a-c))}{b\pi(a-c)} = \frac{\textrm{sinc}(a-c)}{b} \; ,
\end{equation}

\noindent Using these integral solutions, we are free to evaluate the separable $x$ and $y$ terms in the integration of Equation \ref{eqn:intensdist}, ultimately yielding the transmission map:

\begin{equation*}
T(\delta_{0y},\delta_{0x},\delta_{1x}) = \int_{-\infty}^\infty \int_{-\infty}^\infty \frac{1}{2} I_{tot}(x,y) \: dxdy =
\end{equation*}

\begin{align*}
\frac{I_0}{2} \left(\frac{ D^2}{\lambda f}\right)^2 \int_{-\infty}^\infty \int_{-\infty}^\infty      \left| \textrm{sinc}\left( \frac{D(x-\delta_{0x})}{\lambda f} - \frac{Dx_0}{\lambda f}  \right)     \textrm{sinc}\left(  \frac{D(y-\delta_{0y})}{\lambda f} - \frac{Dy_0}{\lambda f}  \right)\right|^2   \\   +    \left| \textrm{sinc}\left( \frac{D(x-\delta_{1x})}{\lambda f} - \frac{Dx_0}{\lambda f}  \right)      \textrm{sinc}\left(  \frac{D(y+\delta_{0y})}{\lambda f} - \frac{Dy_0}{\lambda f}  \right)\right|^2  \\  - 2 \: \textrm{sinc}\left( \frac{D(x-\delta_{0x})}{\lambda f} - \frac{Dx_0}{\lambda f}  \right)     \textrm{sinc}\left(  \frac{D(y-\delta_{0y})}{\lambda f} - \frac{Dy_0}{\lambda f}  \right) \\ \times  \textrm{sinc}\left( \frac{D(x-\delta_{1x})}{\lambda f} - \frac{Dx_0}{\lambda f}  \right)      \textrm{sinc}\left(  \frac{D(y+\delta_{0y})}{\lambda f} - \frac{Dy_0}{\lambda f}  \right)  \: dxdy
\end{align*}

\begin{equation}
\label{eqn:transmission}
= I_0 D^2 \left( 1 -   \textrm{sinc}\left(\frac{2D}{\lambda f}\delta_{0y}\right)   \textrm{sinc}\left(\frac{D}{\lambda f}(\delta_{0x} - \delta_{1x})\right)  \right) \; .
\end{equation}

\begin{figure}
\begin{center}
\begin{tabular}{c}
\includegraphics[height=6cm]{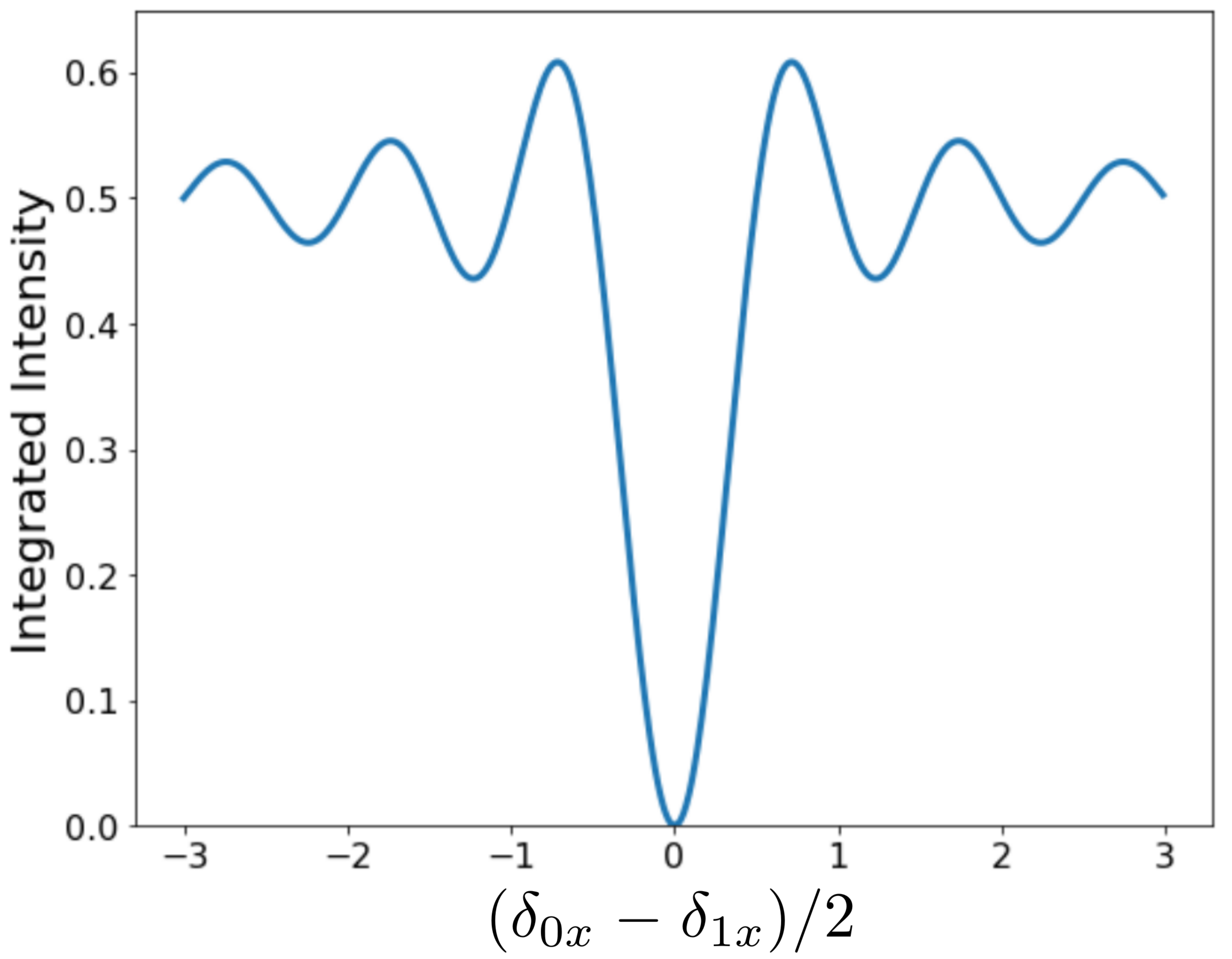}
\end{tabular}
\end{center}
\caption 
{ \label{fig:T}
Total transmission T as a function of $(\delta_{0x} - \delta_{1x})/2$. When the difference of these parameters is zero, corresponding to equal incidence angle on both of the gratings, the single-wavelength PSFs completely overlap and null. For $\delta_{0x} - \delta_{1x} >> 0$, the degree of overlap becomes small, and the total transmission tends to $I_0/2$, as half of the incident intensity is directed toward the `light dump' of the interferometer.
}
\end{figure}

\noindent A factor of $1/2$ is included in the above integral since half of the input light is relayed to the constructive interference path (light dump) of the interferometer. In the last line we have also remembered to include the factor $(\lambda f)^{-2}$ resulting from the $h_0$ term. Equation \ref{eqn:transmission} shows that the transmission function resembles the PSF associated with an aperture of width $2D$ (see Fig. \ref{fig:T}). This result is in line with Section 2.3.3 of the AIC publication\cite{RABBIA2007385}, and demonstrates that DLC also results in $\lambda/2D$ effective angular resolution. However, since the center of the PSF in the x-direction of the focal plane is determined from the grating equation, the $2D$ angular resolution in the secondary telescope focal plane x-direction corresponds to $2L$ effective angular resolution on the sky, where L is the grating length.\cite{tel} The resolution in the focal plane y-direction also exhibits a $2D$ angular resolution, but does not benefit from the resolvance of the grating.

$\delta_{0x},\delta_{1x}$ are focal plane positional shifts arising from the angular offset between a close companion as viewed by Grating 1 and Grating 2 at differing angles of incidence $\alpha$ (Figure \ref{fig:diff}). If the grating is aligned with the angular separation of the close companion and host star ($\beta = 0$), each grating will view the companion on one side or the other of the grating normal. The result is a finite difference between the parameters $\delta_{0x},\delta_{1x}$ ($\delta_{0x} - \delta_{1x} \neq 0$).  As the grating will be used for optical leverage (output angles from the grating are larger than input angles as seen on the sky, which is easily shown via differentiation of the grating equation, with the result  $d \theta_{x}/d \alpha = \textrm{cos}(\alpha)/\textrm{cos}(\theta_{x})$), the angular offset parameters $\delta_{0x},\delta_{1x}$ will generally be much more significant than the oblique offset $\delta_{0y}$. Figure \ref{fig:T} shows the transmission map as a function of $\delta_{0x} - \delta_{1x}$. The physical significance of $\delta_{0x} - \delta_{1x}$ will be further explained in Section \ref{sect:sims}.

\section{\label{sect:sims}Simulated Intensity Distributions}

In the DLC arrangement, each of the two telescopes receives a spectrum from the star and planet being observed. The DLC interferometer nulls the host star by overlapping all single wavelength PSFs in the focal-plane of the combined telescope beams, so that the two beams destructively interfere. The positions in the focal plane of these single wavelength PSFs (of the host star or companion in isolation), in the notation of the previous section, will be determined by:

\begin{equation}
\label{eqn:delta}
\delta_{0x}/f =  \theta_E - \textrm{sin}^{-1}\left(\frac{\lambda}{p} - \alpha\right) \; , \; \delta_{1x}/f = \theta_E -  \textrm{sin}^{-1}\left(\frac{\lambda}{p} + \alpha\right)   \; , \; \delta_{0y}/f = \beta \; ,
\end{equation}

\noindent where $\theta_E$ is the angle between the secondary telescope optical axis and the grating normal (Figure \ref{fig:coords}b), and $p$ is the grating pitch. It is worth noting that Equation \ref{eqn:delta} describes the position of the PSFs of the secondary telescope apertures (in this case a square mirror with side length D). The functional form of these PSFs, as well as the effects of interfering the PSFs from each aperture with each other, were derived in the previous section. The resolution effects of the grating arise from the physical displacements $\delta_{x/y}$, which are determined by the input angles $\alpha$ \& $\beta$, the central exodus angle $\theta_E$, and the ratio $\frac{\lambda}{p}$. $\theta_E$ and $\frac{\lambda}{p}$ are fundamentally tied to the observable grating length L (for more information, see Reference 18). The derivations are done in terms of the secondary aperture PSFs because these are the actual \textit{physical} PSFs landing on the focal plane (because each secondary aperture focuses a plane wave emitted from the grating at each wavelength $\lambda$).

Equation \ref{eqn:delta} assumes that the oblique ($\beta$) angle is unchanged after diffraction from the grating, an assumption of paraxial diffraction theory. Given that all input angles to the grating will typically be extremely small for close-companion coronagraphy, this approximation is reasonable. This equation also assumes a small angle approximation of the form $\textrm{sin}(\alpha) \approx \alpha$ in its derivation from the grating equation.

In the case of zero angular offset in the non-dispersing (oblique) direction ($\beta = 0$), and small finite angular offset in the dispersion ($\alpha$) direction, the angular offset between the single-wavelength wavefronts received by Telescope 1 and Telescope 2 is given by:

\begin{equation}
\label{eqn:dtheta}
\Delta \theta_x  \approx \textrm{sin}^{-1}\left(\frac{\lambda}{p} + \alpha\right) - \textrm{sin}^{-1}\left(\frac{\lambda}{p}  -\alpha\right).
\end{equation}

\noindent It is worth noting that for small values of the angular offset $\alpha$, $\Delta \theta_x$ of Equation \ref{eqn:dtheta} is approximately double the angular offset between the host star and close companion $\delta_{1,0x}$ given in Equation \ref{eqn:delta}. The angular separation between the shifted single-wavelength PSFs is therefore double the angular separation between the close companion and the host star. The close companion is not nulled provided the single-wavelength PSFs do not significantly overlap, so this factor of two increase in the angular separation allows for $\lambda/2D$ effective resolution; this result is similar to the resolution limits attained by AIC\cite{RABBIA2007385}. As mentioned previously, the grating converts this $\lambda/2D$ resolution into a $\lambda/2L$ resolution in the direction of dispersion (focal plane $x$ coordinate), while the resolution in the oblique direction ($y$ coordinate) is simply $\lambda/2D$, since the observed width of the grating is the same as the diameter of the telescope.

Using the functional forms of $I(x,y)$ and $T(\delta_{0y},\delta_{0x},\delta_{1x})$ derived in previous sections, we can now plot the intensity distributions alongside the transmission map.

\begin{figure}[h!]
\begin{center}
\begin{tabular}{c}
\includegraphics[width=0.8\linewidth]{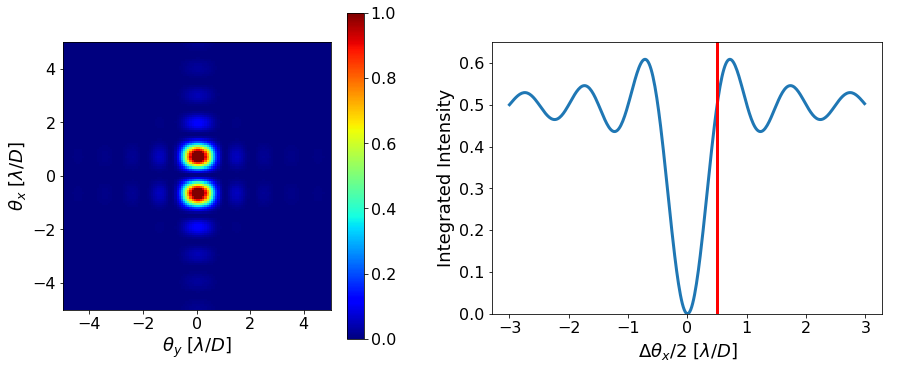}
\end{tabular}
\end{center}
\caption 
{ \label{fig:single}
Intensity distribution of single-wavelength PSFs (left) and transmission function (right) for the case of $\Delta \theta_x/2 = 0.5 [\lambda/D]$. The coordinates of the distribution maps are angular position in terms of resolution elements $\lambda/D$. Note that the coordinates of the transmission map are not equivalent to angular position on the sky, as would be expected in a conventional telescope. Instead, sky angular position ($\alpha , \beta$) must be determined from Equations \ref{eqn:delta} \& \ref{eqn:dtheta}. The right plot illustrates the total integrated intensity in the focal plane as a function of angular offset of the source in the focal plane. The red line indicates the angular offset for the intensity distribution on the left. Note that this is the distribution of a single object/input-wave, with the two PSFs corresponding to differing incidence angle on Grating 1 vs. Grating 2. The special case of $\Delta \theta_x/2 = 0.5 [\lambda/D]$ is chosen as it represents an angular separation of half the typical resolution limit of the gratings. The color indicates the transmission fraction. Significant transmission at $\Delta \theta_x/2 = 0.5 [\lambda/D]$ indicates a factor 2 improvement over diffraction-limited imaging. 
}
\end{figure} 

\pagebreak

Figure \ref{fig:single} gives the intensity distribution for a single angular offset corresponding to $\Delta \theta_x/2 = 0.5 [\lambda/D]$, the angular separation corresponding to half the diffraction limit of the gratings. It is seen that there is significant transmission at this angular separation, allowing for observation of any close-companion with angular separation greater than (or equal to) this condition.

\begin{figure}[h]
\begin{center}
\begin{tabular}{c}
\includegraphics[height=6.1cm]{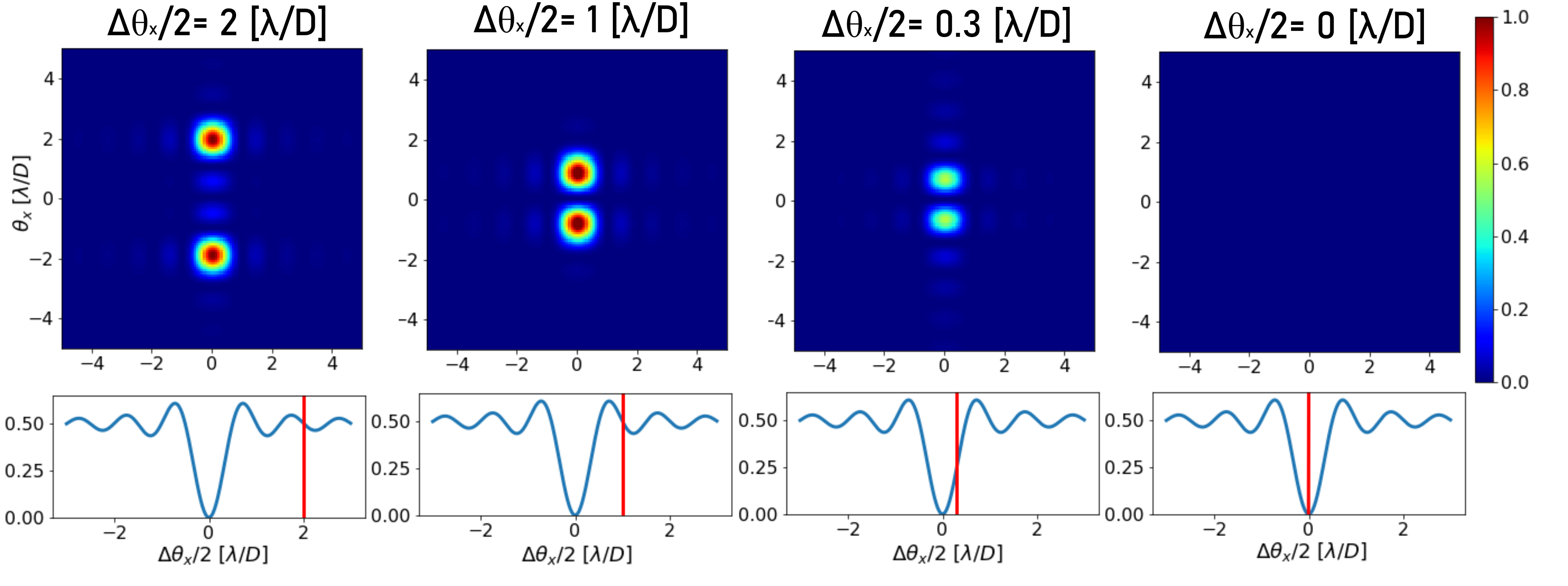}
\end{tabular}
\end{center}
\caption 
{ \label{fig:singlemultiple}
Four cases of steadily decreasing $\alpha$ (with corresponding decrease in ($\delta_{0x}-\delta_{1x}$) and $\Delta \theta_{x}$). Color in the upper row of plots indicates the transmission fraction. The location on the transmission map is displayed beneath each intensity distribution. As $\alpha$ decreases, corresponding to increasingly normal incidence on the gratings, the single-wavelength PSFs start to merge and ultimately null. The rightmost column illustrates ideal nulling, in which the PSFs perfectly overlap everywhere in the focal plane with zero phase errors in the optical train.
}
\end{figure}

As the angular separation between companion and host-star ($\alpha$) decreases, there is a corresponding decrease in $\Delta \theta_{x}$ as described by Equation \ref{eqn:dtheta}. Decreasing $\alpha$ therefore pushes single-wavelength PSFs closer together in the focal plane, until they ultimately begin to overlap and increasingly null. A visualization of this phenomenon is shown in Figure \ref{fig:singlemultiple}.

\pagebreak

\subsection{\label{sect:poly}Full-Bandwidth, Polychromatic Intensity Distributions}

All of the preceding discussion has focused on the case of a single wavelength (a monochromatic input wave). In reality, there will be many wavelengths incident on the grating, with a slice of the total spectrum entering the secondary telescope after diffraction from the grating. In the POG arrangement, the location in the focal plane $x$ coordinate is correlated with wavelength. As such, the nulling processes previously discussed will occur at a different location in the focal plane for each wavelength in the spectrum that is received by the secondary telescope. Each wavelength will be nulled at the location in the focal plane corresponding to zero angular offset in $\alpha$ and $\beta$. For a broadband source at normal incidence to the gratings, all single wavelength PSFs will fully overlap and null in an ideal scenario. 

\begin{figure}[h]
\begin{center}
\begin{tabular}{c}
\includegraphics[width=1\linewidth]{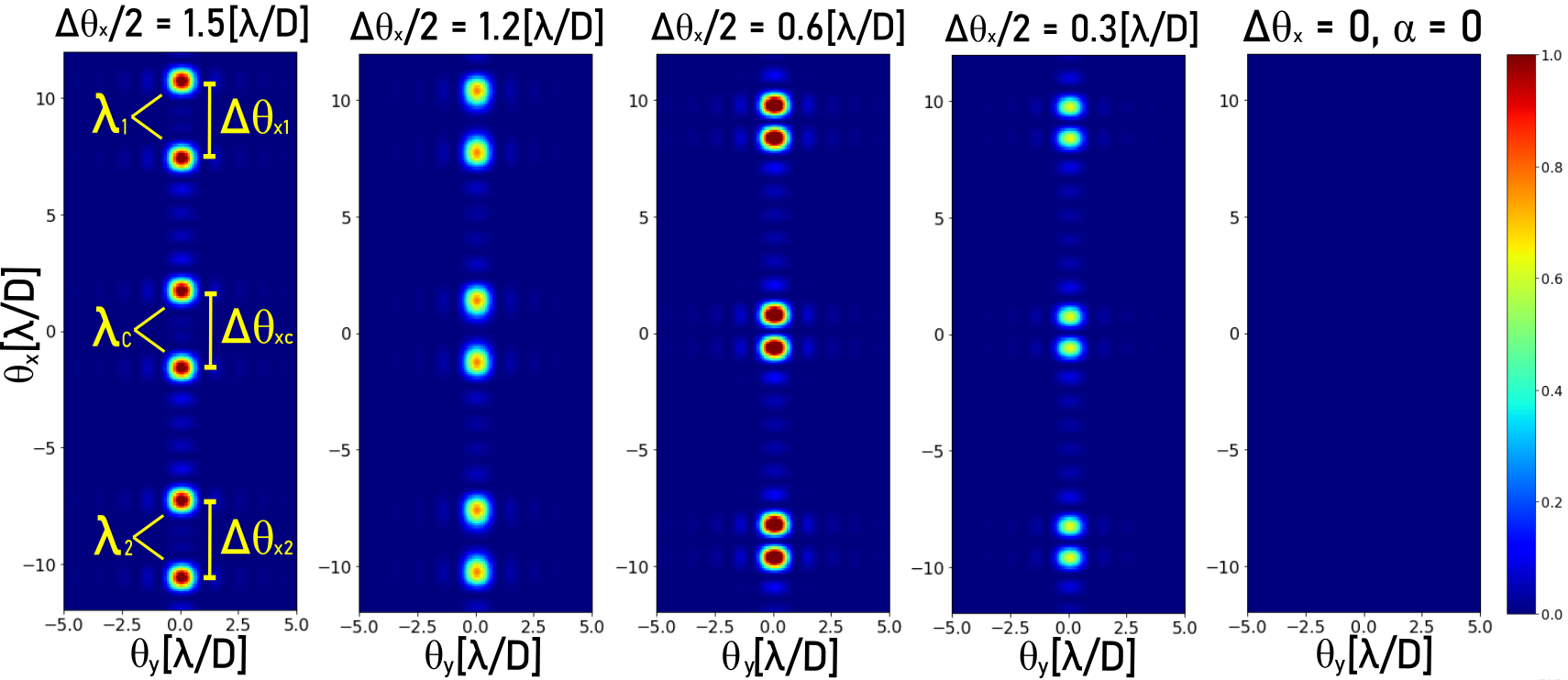}
\end{tabular}
\end{center}
\caption 
{ \label{fig:multiple}
Intensity distributions for the case of a single object with a spectrum composed of three very narrow bands. The coordinates of the distribution maps are angular position in terms of resolution elements $\lambda_c/D$. Angular position in the focal plane is not equal to angle on the sky, as would be the case for a conventional telescope. Instead, sky angular position must be determined from Equations \ref{eqn:delta} \& \ref{eqn:dtheta}.  $\lambda_c$ is the central wavelength of the stellar/planet spectrum, received by the secondary telescope with zero angular offset from the optical axis of the secondary aperture. Moving from left to right, the angle of incidence to the grating (angle on the sky $\alpha$) is decreased, with a corresponding decrease in $\Delta \theta_x$. At $\alpha = 0$ \& $\Delta \theta_x = 0$, the PSF pairs of all three wavelengths merge together and are nulled. Variations in brightness during the merger are explained by the angular positions of each PSF pair on the single-wavelength transmission map (as shown in the bottom row of Figure \ref{fig:singlemultiple}). For a full-bandwidth scenario, one can imagine an infinite number of PSF pairs, constituting a complete spectrum from each of the two secondary telescopes, shifted by an angular difference $\Delta \theta_x$. When a broadband source is received at normal incidence to the gratings, the two spectra overlap with zero angular offset and are therefore nulled everywhere.
}
\end{figure}

Figure \ref{fig:multiple} demonstrates polychromatic nulling for three distinct wavelengths for an object (such as an exoplanet) that is located along the long axis of the grating ($\beta=0$), but has a small angular distance from the grating normal (small $\alpha$). This figure demonstrates that the focal plane will record a small piece of the spectrum of each object that is within the field of view of the secondary telescope in the $\beta$ direction, and within the detectable wavelength range in the $\alpha$ direction. The small piece of the spectrum will be spread out over the entire focal plane in the $\theta_x$ direction. Light from objects with the same $\beta$ in the sky will land on the same $\theta_y$ in the focal plane and completely overlap each other. However, the wavelengths will be shifted by an amount that depends on the angle $\alpha$ from the normal. When $\alpha$ is small (within the diffraction limit), the entire detected wavelength range from that object will be nulled.  

\pagebreak

\section{\label{sect:perf}DLC Performance Considerations}

\subsection{\label{sect:stelleak}Stellar Leakage}

\begin{figure}[h!]
\begin{center}
\begin{tabular}{c}
\includegraphics[width=0.3\linewidth]{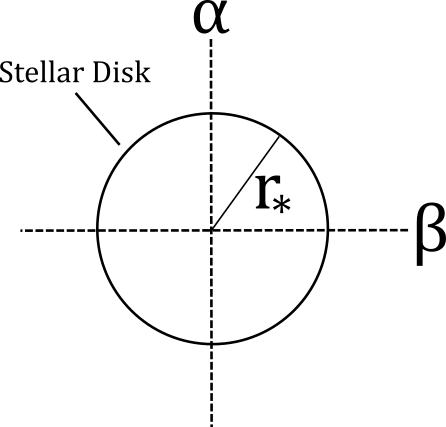}
\end{tabular}
\end{center}
\caption 
{ \label{fig:disk}
Finite size of the stellar disk in on-sky ($\alpha$, $\beta$) coordinates.
}
\end{figure}

Stellar leakage, occurring as a result of the finite angular extent of the stellar disk (Figure \ref{fig:disk}), limits performance in virtually all nulling architectures. The amount of stellar leakage $\omega$ will be determined by integrating the curve of Figure \ref{fig:T} over the minuscule angular range within the stellar radius $r_*$, centered at zero. This calculation assumes a stellar disk of uniform brightness distribution.

\begin{equation}
\label{eqn:wabt}
\omega(\alpha, \beta) =  \frac{4}{\pi r_*^2}\int_{0}^{r_*} \int_{0}^{\sqrt{{r_*}^2 - \beta^2}} T(\alpha,\beta) \: d\alpha d\beta
\end{equation}

\noindent In previous sections, the transmission map $T$ was given in terms of focal plane coordinates $x$ and $y$. For the purposes of this analysis, the quantities $\delta_{0x}$, $\delta_{1x}$, $\delta_{0y}$ previously taken to be constant in the derivation of $T$, will be given explicit $\alpha$ $\beta$ dependence according to Equation \ref{eqn:delta}. Plugging the transmission into Equation \ref{eqn:wabt}, we find:

\begin{equation}
\label{eqn:omega}
\omega(\alpha, \beta) =  \frac{4 I_0 D^2}{\pi r_*^2} \int_{0}^{r_*} \int_{0}^{\sqrt{{r_*}^2 - \beta^2}} \scriptstyle \left[ 1 -   \textrm{sinc}\left(\frac{2D}{\lambda }\beta \right)      \textrm{sinc}\left(\frac{D}{\lambda }\left(\textrm{sin}^{-1}\left(\frac{\lambda}{p} + \alpha\right) - \textrm{sin}^{-1}\left(\frac{\lambda}{p} - \alpha\right)\right)\right)  \right]  \: \displaystyle d\alpha d\beta.
\end{equation}

\noindent To evaluate the $\alpha$ integral, we will approximate the following term as a Taylor series:

\begin{equation}
\label{eqn:taylor}
\textrm{sinc}\left(\frac{D}{\lambda }\left(\textrm{sin}^{-1}\left(\frac{\lambda}{p} + \alpha\right) - \textrm{sin}^{-1}\left(\frac{\lambda}{p} - \alpha\right) \right)\right) \approx 1 \; - \; \alpha^2 \frac{2}{3}\left( \frac{D\pi}{\lambda}\right)^2(1-\lambda^2/p^2)^{-1}
\end{equation}

\noindent In practice the angular extent of the stellar disk $r_*$ (and correspondingly the domain of $\alpha$) will be very small, so contributions from higher terms of the series should have minor impact on the stellar leakage estimate. Plugging this expression into Equation \ref{eqn:omega} yields:

\begin{equation}
\label{eqn:bint}
\omega(\alpha, \beta) \approx \frac{4 I_0 D^2}{\pi r_*^2} \int_{0}^{r_*}  \scriptstyle \left(\sqrt{{r_*}^2 - \beta^2} \; - \; \textrm{sinc}\left(\frac{2D}{\lambda }\beta \right)   \left(\sqrt{{r_*}^2 - \beta^2}  -  ({r_*}^2 - \beta^2)^{3/2}\frac{2}{9}\left( \frac{D\pi}{\lambda}\right)^2(1-\lambda^2/p^2)^{-1}\right) \right) \displaystyle d\beta.
\end{equation}

At this point we will employ a second Taylor series approximation in order to simplify the integral:

\begin{equation}
\textrm{sinc}\left(\frac{2D}{\lambda }\beta \right) \approx 1 - \frac{2\pi^2D^2}{3\lambda^2}\beta^2 
\end{equation}

\noindent After applying this approximation to the integral of Equation \ref{eqn:bint}, the stellar leakage is found to be:

\begin{equation}
\label{eqn:leaky}
\omega(r_*) \approx I_0 r_*^2  \frac{D^4\pi^2}{6\lambda^2} \left( 1 + (1 - \lambda^2/p^2)^{-1} - r_*^2 \: \left( \frac{D\pi}{3\lambda} \right)^2 (1-\lambda^2/p^2)^{-1} \right).
\end{equation}

At a distance of 10 parsecs, a Sun-like star will have an angular radius of $\sim 2.5 \times 10^{-9}$ radians. Assuming current DICER benchmarks of a 1m secondary telescope, 10$\mu$m central wavelength, and a grating pitch of 10.055$\mu$m (resulting from a $84^\circ$ exodus angle), Equation \ref{eqn:leaky} yields a stellar leakage estimate of $\omega \approx 9.5 \times 10^{-6} I_0 D^2$ ($I_0 D^2$ is the power received under total transmission). In this scenario, flux from the star cannot be reduced/rejected beyond a factor of $\sim 10^5$ ($1/9.5 \times 10^{-6}$).

\subsection{\label{sect:jitleak}Telescope Pointing Error/Jitter}

In a manner analogous to AIC and Bracewell configurations, the effectiveness of DLC is contingent on the ability of the telescope to point accurately in the direction of the host star. The following analysis is similar to that found in Rabbia et al., but is unique to POG architectures. We define the level of rejection (i.e., factor by which the host-star flux is reduced) as:

\begin{equation}
\label{eqn:rej}
\left< Rej \right> = \frac{T_0}{\left<T(\alpha, \beta) \right>} ,
\end{equation}

\noindent where $T_0$ is taken to be the case of total transmission (no nulling), and $\left<T(\alpha, \beta) \right>$ is the average transmission under conditions of randomly varying $\alpha, \beta$, due to telescope pointing error/jitter. After cancellation of the shared constants, equation \ref{eqn:rej} can be written:

\begin{equation}
\label{eqn:jittone}
\left< Rej \right>  = \frac{1}{\left< 1 - \textrm{sinc}(\frac{2D}{\lambda}\beta)\textrm{sinc}\left(\frac{D}{\lambda }\left(\textrm{sin}^{-1}\left(\frac{\lambda}{p} + \alpha\right) - \textrm{sin}^{-1}\left(\frac{\lambda}{p} - \alpha\right) \right)\right) \right>}.
\end{equation}

\noindent Utilizing the Taylor series approximation of Equation \ref{eqn:taylor}, and applying yet another series approximation of $\textrm{sinc}(x) = 1 - \frac{\pi^2x^2}{6}$, we obtain:

\begin{equation}
\label{eqn:rejfin}
\left< Rej \right> = \frac{1}{\left< \alpha^2 \right>\frac{2\pi^2 D^2p^2}{3\lambda^2p^2-3\lambda^4} + \left< \beta^2 \right> \frac{2\pi^2 D^2}{3\lambda^2}}.
\end{equation}

In the above expression, cross terms containing $\alpha^2*\beta^2$ have been omitted because $\alpha$\&$ \beta$ are small. Reading Equation \ref{eqn:rejfin}, it is clear that DLC will respond differently to pointing error in the on-sky coordinates $\alpha$ and $ \beta$, with more stringent requirements placed on $\alpha$ than $\beta$ in order to achieve a target rejection level. This asymmetrical angular dependence may inform the design of fine-guidance systems for DLC applications. 

In order to resolve an Earth-like exoplanet from a Sun-like star at 10$\mu$m, we aim to null the host star by as much as $10^5$ with the DLC. In DLC, pointing jitter is likely the most important factor for achieving high degrees of nulling (as is the case with conventional AIC). If we assume a secondary aperture width of 1m, a wavelength of 10$\mu$m, and a grating pitch of 10.055$\mu$m (resulting from a $84^\circ$ exodus angle), Equation \ref{eqn:rejfin} shows that an RMS pointing jitter of $\sigma_\alpha = \sqrt{\left< \alpha^2 \right>} = 1.25\times10^{-9}$ rad and $\sigma_\beta = 2.5\times10^{-9}$ rad will result in $\left< Rej \right> \approx 10^5$. The RMS pointing jitter of JWST is approximately 1mas, or $\sim 5\times10^{-9}$ radians. DICER may therefore require fine-guidance capability that is four times more precise than JWST (for the $\alpha$ axis) if a factor $10^5$ nulling is desired. However, if limitations imposed by the local zodiacal dust background and stellar leakage render $10^5$ nulling irrelevant, we could simply choose to null by a factor of $10^4$ with DICER. This degree of nulling is achievable for $\sigma_\alpha = \sigma_\beta  \approx 0.8\textrm{mas}$ (similar to the fine guidance capability of JWST).

\subsection{Broadband Correction to Leakage and Jitter}

The calculations for stellar leakage and pointing jitter given in sections \ref{sect:stelleak} and \ref{sect:jitleak} are valid only at a single wavelength $\lambda$. In this section we will show that for the DICER application the broadband calculations change the results very little. The derived equations might be necessary, however, for applications that transmit a wider wavelength range.

In the example calculations for DICER, $\lambda$ was chosen to be $\lambda_c = 10 \mu$m, the center of the DICER observation bandwidth. However, both the stellar leakage and pointing jitter will exhibit a spectral response resulting in variation across the observation bandwidth. In order to give a more accurate estimation of leakage and jitter, we must integrate expressions \ref{eqn:leaky} and \ref{eqn:rejfin} over the observation bandwidth $\Delta \lambda$.

For the stellar leakage, we obtain:

\begin{equation}
\label{eqn:leaky_int}
\int^{\lambda_c + \Delta \lambda/2}_{\lambda_c - \Delta \lambda/2} \omega \; d \lambda  = \frac{I_0r_*^2D^4\pi^2}{6} \times
\end{equation}
\begin{equation*}
 \left[ \left( \left( \frac{r_*D\pi}{3p} \right)^2 -2\right)\lambda^{-1} + \left( \frac{r_*D\pi}{3} \right)^2 \frac{\lambda^{-3}}{3} + \frac{\textrm{tanh}^{-1}(\lambda/p)}{p}\left( 1- \left( \frac{r_*D\pi}{3p} \right)^2\right)  \right]
\Biggr|_{\lambda_c - \Delta \lambda/2}^{\lambda_c + \Delta \lambda/2}.
\end{equation*}

\noindent Dividing the above expression by  $\Delta \lambda$ (to take the average leakage in band) and plugging in the DICER bandwidth of $\Delta \lambda = 14.7$nm with $\lambda_c = 10\mu$m, yields a stellar leakage of $\omega \approx 9.6 \times 10^{-6} I_0 D^2$. This result is very close to the value found in Section \ref{sect:stelleak}, which is to be expected since the DICER bandwidth is very small.

For the wavelength response of the pointing jitter, we will focus on the $\alpha$ axis, as this was shown in Section \ref{sect:jitleak} to have more sensitive requirements than the $\beta$ axis. This assumption will also facilitate a tilt compensation calculation which will be applied later under the assumption of $\beta = 0$. To calculate the average broadband transmitted fraction of light, we integrate the transmission fraction, $<Rej>^{-1}$, over the wavelength band and then divide by the width of the band:

\begin{equation}
\label{eqn:leaky_int}
\frac{1}{\Delta \lambda}\int^{\lambda_c + \Delta \lambda/2}_{\lambda_c - \Delta \lambda/2} \left< Rej \right>^{-1}\biggr|_{\beta=0} \; d \lambda = \frac{1}{\Delta \lambda} 
\left(- \frac{\left<\alpha^2 \right>2\pi^2D^2p^2\left(p-\lambda\textrm{tanh}^{-1}(\lambda/p)\right)}{3p^3\lambda} \right)\Biggr|_{\lambda_c - \Delta \lambda/2}^{\lambda_c + \Delta \lambda/2}.
\end{equation}

Applying the DICER bandwidth of $\Delta \lambda = 14.7$nm with $\lambda_c = 10\mu$m, an average pointing jitter of $\sigma_\alpha = \sqrt{\left< \alpha^2 \right>} = 1$mas yields an average rejection factor of $\sim7000$ across the band. This is slightly worse than the single wavelength result given in section \ref{sect:jitleak}, demonstrating that there is some degradation of the nulling at the edges of the bandwidth.

The nulling degradation becomes more pronounced for larger jitter values. For example, if we repeat the above calculation using $\sigma_\alpha = \sqrt{\left< \alpha^2 \right>} = 10$mas, we achieve a nulling factor $\left< Rej \right> \approx 70$. 10mas jitter was not arbitrarily chosen, being approximately equal to the pointing stability of JWST in the absence of tilt corrections. 

We will now apply such a tilt correction to the derivation of Equation \ref{eqn:leaky_int}, to show that the nulling can in principle be recovered. Starting with Equation \ref{eqn:jittone}, we can introduce an achromatic tilt correction factor $C = \textrm{sin}^{-1}\left(\frac{\lambda_c}{p} + \alpha\right) - \textrm{sin}^{-1}\left(\frac{\lambda_c}{p} - \alpha\ \right)$:

\begin{equation}
\left< Rej_c \right>\biggr|_{\beta=0 } = \frac{1}{\left< 1 - \textrm{sinc}\left(\frac{D}{\lambda }\left(\textrm{sin}^{-1}\left(\frac{\lambda}{p} + \alpha\right) - \textrm{sin}^{-1}\left(\frac{\lambda}{p} - \alpha\right) - C \right)\right) \right>}.
\end{equation}

\noindent The correction term is constructed such that the PSFs from each grating perfectly overlap and null at the central wavelength $\lambda_c$. However, the chromatic degradation found earlier will not be perfectly canceled by this tilt correction. Once again applying the Taylor series approximation of Equation \ref{eqn:taylor}, we may integrate the inverse rejection factor as in Equation \ref{eqn:leaky_int}:

\begin{equation}
\int^{\lambda_c + \Delta \lambda/2}_{\lambda_c - \Delta \lambda/2} \left< Rej_c \right>^{-1}\biggr|_{\beta=0} \; d \lambda = \int^{\lambda_c + \Delta \lambda/2}_{\lambda_c - \Delta \lambda/2} \frac{ \left<\alpha^2 \right> D^2 \left[ 2(1-\lambda^2/p^2)^{-1/2} - 2(1-\lambda_c^2/p^2)^{-1/2} \right]^2 }{6\lambda^2} \; d \lambda
\end{equation}

\begin{equation*}
= \frac{\left<\alpha^2\right>D^2}{6\lambda}\left( 4\sqrt{1-\frac{\lambda_c^2}{p^2}}\sqrt{1-\frac{\lambda^2}{p^2}} + \frac{4\lambda\textrm{tan}^{-1}(\lambda/p)}{p} + \frac{\lambda_c^2}{p^2} -5 \right)\Biggr|_{\lambda_c - \Delta \lambda/2}^{\lambda_c + \Delta \lambda/2} .
\end{equation*}

\noindent The above equation gives the tilt-corrected broadband pointing jitter response of DLC. Dividing by $\Delta \lambda = 14.7$nm and plugging in the previously used values of $p$, $\lambda_c$, $D$, and $\sigma_\alpha = \sqrt{\left< \alpha^2 \right>} = 10$mas yields an average residual starlight of $7.8 \times 10^{-6} I_0D^2$, or a rejection factor $\sim10^5$, thus recovering the previously achieved rejection. We see that even though there is degradation of the broadband jitter response, it is effectively compensated by an (ideal) tilt correction term.

\subsection{Residual Optical Path Difference}

In the preceding analyses, new derivations of DLC performance characteristics were required due to the vastly different geometry of the DLC configuration (compared with AIC). However, in the case of Optical Path Difference (OPD) leakage, the situation is unchanged by the asymmetrical and chromatic nature of the dispersive primary. Since DLC in the single-wavelength view is functionally identical to AIC, we take the rejection limit under conditions of non-zero OPD to be equal to that found in Rabbia et al.\cite{RABBIA2007385} :

\begin{equation}
Rej = \frac{\lambda^2}{\pi^2d^2},
\end{equation}

\noindent where $d$ is taken to be either residual OPD or the average of surface figure error. The above expression may be rearranged to yield the condition:

\begin{equation}
d \le \frac{\lambda}{\pi\sqrt{Rej}},
\end{equation}

\noindent where here $Rej$ is the target rejection. As an axample, if a factor of $10^4$ nulling were desired, the residual OPD would need to be reduced (through active feedback adjustments) to $d\approx \lambda/100\pi$.

\subsection{Bandwidth Invariance for a Fixed Resolution}

It was shown in a previous publication\cite{tel} that bandwidth strictly decreases with increasing exodus angle ($\theta_E$ in figure \ref{fig:diff}b). However, this result was found for a system in which the central wavelength $\lambda_c$ (the wavelength received at the angle $\theta_E$) is held constant as $\theta_E$ is increased, with all other grating parameters adjusted to ensure this condition. 

When considering a form-factor for a DLC arrangement requiring a target angular resolution $\gamma$ (e.g. 0.1 arcseconds, the angular separation of the Earth-Sun system at 10pc distance), it makes more sense to hold $\gamma$ constant, while varying all other telescope parameters. Under these conditions, the resulting optical system exhibits the counter-intuitive result that the overall bandwidth $\Delta \lambda$ is invariant with respect to $\theta_E$. This result can be demonstrated mathematically as follows. 

For a POG telescope, bandwidth is defined as the difference between the wavelengths received at the edges of the secondary FOV:

\begin{equation}
\Delta \lambda = p \left( \textrm{sin}(\theta_E+ \textrm{FOV/2}) - \textrm{sin}(\theta_E- \textrm{FOV/2}) \right) \; .
\end{equation}

\noindent Making the substitution $p = \lambda_c/\textrm{sin}(\theta_E)$ and utilizing the compound angle formula results in a simplified expression:

\begin{equation}
\Delta \lambda =  \frac{2\lambda_c}{\textrm{tan}(\theta_E)}\textrm{sin}(\textrm{FOV}/2) \; .
\end{equation}

\noindent The invariance of $\Delta \lambda$ is then demonstrated by making the substitution $\theta_E \approx \pi/2 - \textrm{tan}^{-1}\left( D/L \right)$, in addition to defining the (constant) target angular resolution to be $\gamma = \lambda_c/2L$, where the factor of two in the denominator is a result of the resolution enhancement of DLC. Plugging into the prior expression yields:

\begin{equation}
\Delta \lambda \approx 4\gamma D \textrm{sin}(\textrm{FOV}/2) \; .
\end{equation}

\noindent The above expression is noteworthy in that it does not depend on any of the interrelated parameters $\theta_E$, $L$, or $ \lambda_c$. This indicates that if one requires a fixed angular resolution in the direction of the long axis of the grating, then the bandwidth depends only upon the choice of the secondary telescope parameters, and is not dependent on any characteristics of the diffraction gratings, or even the wavelength at which one chooses to observe.

\section{Finding and Measuring Exoplanets with DICER}

Here, we work out an example application (DICER) of the DLC that aims to find nearby (within 8 pc) habitable exoplanets. As we will see, DICER can in principle find nearby, habitable exoplanets. However, because exoplanets are very faint compared to the zodiacal and exozodiacal light backgrounds, DICER requires a very complex secondary spectrograph system that can disentangle the wavelength and sky position of the emitting object for photons in the focal plane; without a second disperser, light from objects whose sky positions differ only in the $\alpha$ direction (aligned with the long axes of the gratings) will lie right on top of each other but will be received at different wavelengths. As we will see, building a second disperser with the required resolution is quite difficult and makes it unlikely that DICER is the best application of the DLC. It does, however, demonstrate the throughput and design requirements of DLC applications.

Imagine pointing DICER, as depicted in Figures \ref{fig:DICER}, \ref{fig:pog}, and \ref{fig:DLCraytrace}, at a nearby star that might or might not host an exoplanet. We imagine a DLC with two 10m by 1m gratings, each with a pitch of 10.055$\mu$m, and two 1m-by-1m infrared telescopes that capture the light that is diffracted from the gratings at a $84^\circ$ exodus angle. In order to collect light in a 14.7 nm-wide passband around 10$\mu$m, the infrared telescopes have to provide a diffraction-limited point spread function over a 0.8$^\circ$ field of view.

\subsection{The Need for Dual Dispersion}

If there is an exoplanet orbiting the star, then the DLC signal in the focal plane consists of two very small and typically overlapping segments of the exoplanet emission spectrum, one from each diffraction grating, in addition to any leaked light from the host star. For perfect point sources, the width of each spectrum covers 2" on the sky, which is the diffraction limit of the 1m width of the grating for 10$\mu$m light. If the planet is closer than 1" to the host star, and a line from the star to the planet is aligned with the short direction of the grating, then the planet light from the two sides of the interferometer will arrive at the focal plane out of phase and right on top of each other, so the exoplanet light will be nulled along with the host star light. The factor 2 difference between the size of the nulled region on the sky and the width of the point spread function (PSF) for point sources is a feature of both of DLC and AIC coronagraphs.

In the long (dispersion) direction of the grating, light from one position in the sky will be spread out across the focal plane; the position where a photon lands depends on both the sky position of the object that emitted the photon and its wavelength. If we measured the wavelength of the light (for example with a spectrograph), we could in principle locate the object in the sky with 0.2" accuracy. To achieve this spatial accuracy, though, would require a spectrograph with the resolution of the primary objective gratings, which for a 10 meter grating with 10$\mu$m pitch translates to R=1,000,000. Fortunately, the null of the coronagraph eliminates a region of the sky that is $0.1" \times 1"$ around the host star, without the need for a second spectrograph.

While ray tracing shows that the host star is nulled to machine precision (in our case by a factor of $10^{14}$), real issues such as stellar leakage (from the nonzero physical size of the host star), optical path difference, and pointing jitter will cause the host star to spill out of the nulled region and therefore contribute to the exoplanet background. Section \ref{sect:perf} showed that stellar leakage limits the suppression of starlight to a factor of $10^5$, calculated for a G star at a distance of 10 pc. With pointing accuracy similar to JWST, the nulling for the same star would be limited to $10^4$. The amount of stellar leakage, and required pointing accuracy, depend on the distance and physical size of the host stars that are observed.

Because the photon rate from an exoplanet is very low, we do not expect to measure fine features of the observed spectrum or the offset in wavelength between the light that comes from the two different gratings. Therefore, for this application we choose to measure all of the photons from both sides of DICER together as one measurement, including the residual light from the host star and all other background sources.

\begin{figure}[h!]
\begin{center}
\includegraphics[width=8.5cm]{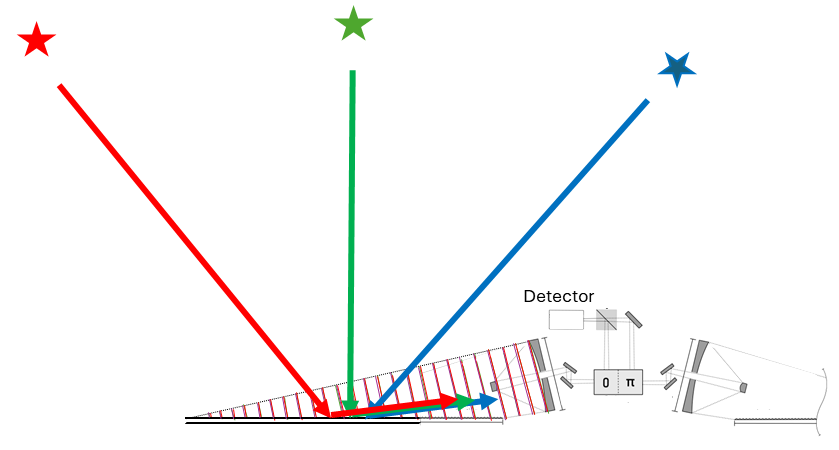}
\end{center}
\caption{Light from a wide area of sky captured by DICER. In the dispersion direction of the grating, DICER can accept light from a large angle of the sky; different wavelength photons from objects at wide angular separation can be diffracted into the same secondary telescope incident angle, and thus land at exactly the same position in the focal plane. Placing a filter in the optical path will limit somewhat the range of wavelengths (and therefore the region of the sky) observed. DICER requires an R=100,000 spectrograph to disambiguate the position and wavelength of the light in the focal plane in order to achieve a resolution of 2" in the dispersion direction of the grating.}\label{fig:skyangle}
\end{figure}

For a {\it single wavelength} point source, the gratings, secondary telescopes, the DLC produce in the focal plane a diffraction-limited spot size of 0.2” in the grating dispersion direction and 2” in the perpendicular direction (note that the limit to resolve the planet from the host star is half the PSF spot size). However, each star in the sky produces a continuum of wavelengths across the entire spectral range at which the focal plane detectors are sensitive. That means that objects at wildly different sky positions (say 60$^\circ$ apart along the grating direction) can produce photons with different wavelengths at the same place in the focal plane (Figure~\ref{fig:skyangle}). Observing all of the light from a part of the sky that is 2” wide and tens of degrees long along with the exoplanet signal would produce an unacceptable background level for finding exoplanets.

The obvious solution is to introduce a filter (anywhere in the optical path where there is a collimated beam) that only transmits the light in a 14.7 nm bandpass, centered on 10$\mu$m. That is the range of wavelengths we expect to detect for objects that are very close to perpendicular to the gratings. While we do expect to use such a filter in the optical path to limit the range of wavelengths that enter the DLC, even a bandpass filter that is this narrow still allows background light from an area of the sky that is approximately 2” by 100” to reach the area of the detector that we will be summing together to measure the signal from the exoplanet. Because the zodiacal light from our Solar System is bright compared to the Earth as viewed from a distance of 10 pc, the background even with this narrow filtering is still too large.

We have considered two possible solutions to this problem. The first possibility is to use an extremely narrow (of order an Angstrom wide) bandpass filter that transmits different wavelengths as a function position in the focal plane. The transmitted wavelength range would need to vary so that it only lets through the light from a portion of the sky within 2” of the host star. Note that if you know the position of the host star in the sky, then you know where in the focal plane light from each wavelength in its spectrum will hit. If the extremely narrow, variable filter is technologically feasible then it would be the preferred solution because it is a relatively simple optical component. However, it is unclear whether this is technologically feasible at 10$\mu$m because infrared capabilities are classified.

The second possibility is to use a high resolution spectrograph that determines with great accuracy the wavelength of every photon, in addition to the position where it hits the focal plane. In principle, this could reduce the region over which the background overlaps the emission from the exoplanet to a spot size of 0.2” $\times$ 2" for each of the two DLC exoplanet images. However, this would require a spectrograph with the same resolution as the primary objective grating, $R=10^6$. We do not know how to create such a resolution in this space application.

Instead, we imagined a set of five $R=100,000$ immersion gratings that mimic those produced for the deselected Space Infrared Telescope for Cosmology and Astrophysics \cite{2016SPIE.9904E..2IK} (SPICA). These immersion gratings were cut from single crystals of ZnSe that were 60mm in length and produce R=30,000 at 10$\mu$m. ZnSe is highly transmissive ($\sim$100\% transmission) at 10$\mu$m, according to previous SPICA publications\cite{2008SPIE.7010E..32K}. By making immersion gratings that are 20 cm in length (slightly more than three times larger than those proposed for the high-resolution SPICA infrared spectrograph), we  could reach $R=100,000$. 

\begin{figure}[H]
\begin{center}
\includegraphics[width=0.9\linewidth]{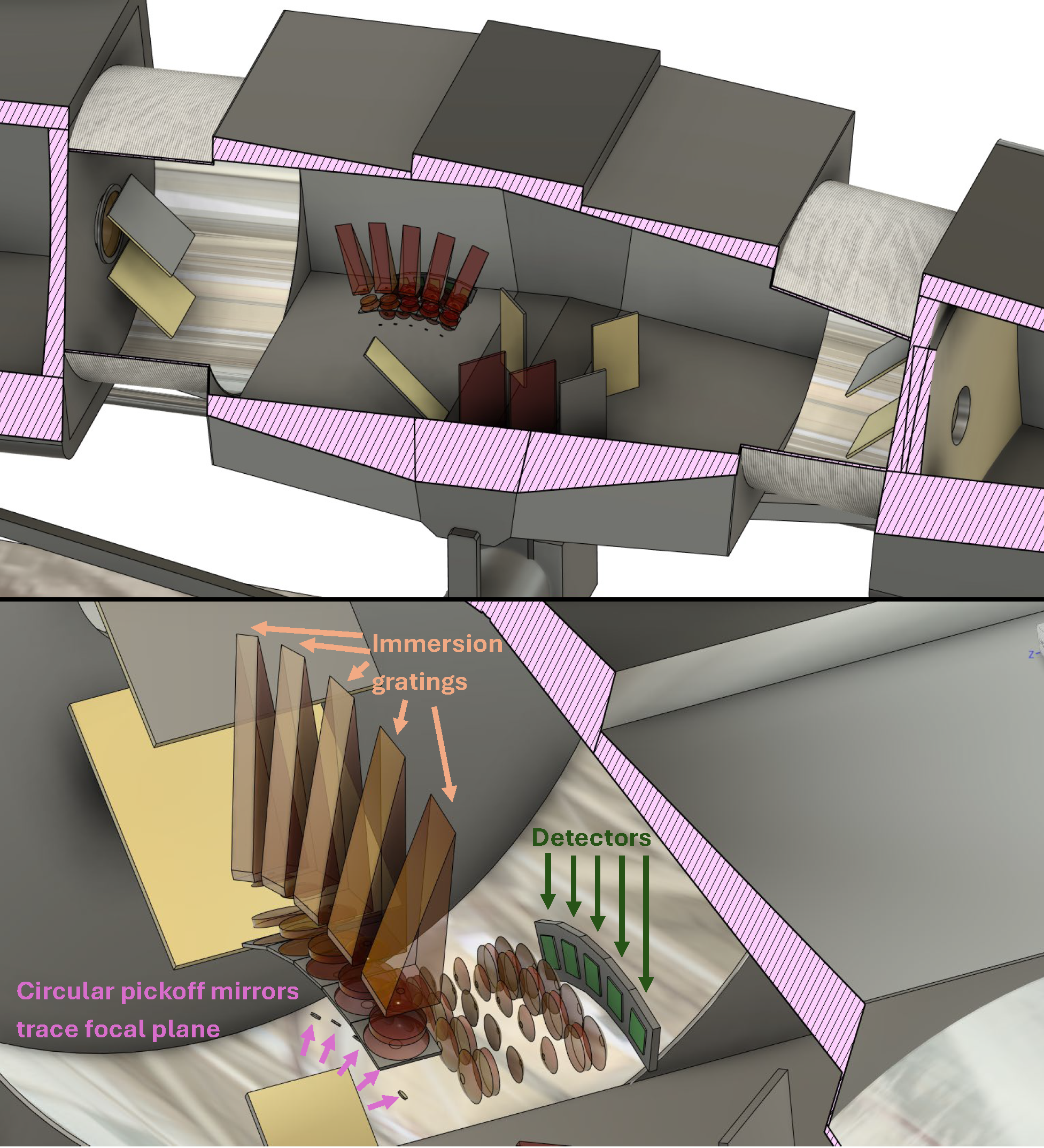}
\end{center}
\caption{CAD drawing of DICER dual dispersion spectrograph. To reduce the region of sky over which zodiacal light background is coincident with the exoplanet signal, it is necessary to disambiguate position and wavelength in the DLC focal plane. In this drawing, the DLC focal plane is traced by the five circular pickoff mirrors located just below the five immersion gratings. Focal plane light that is not nulled passes through focus and then up into the immersion gratings. Five separate spectrographs collect all of the light from the long, curved focal plane. The immersion gratings are single crystals of ZnSe that are 20cm in length, and carved into right rectangular prisms with a grating ruled into the largest, rectangular side. At 10$\mu$m the gratings can achieve $R \sim 100,000$. Light is diffracted off the crystal side of the grating and then the light goes back down and out through the bottom surface of the crystal and is focused onto five pickoff mirrors in the focal plane. It is then corrected and focused onto five detectors, shown in green.}\label{fig:dicer_inside_3}
\end{figure}

For the spectrograph design, we rely on the years of development that have already been done towards SPICA's MIR spectrograph. Figure~\ref{fig:dicer_inside_3} shows a CAD drawing of the SPICA spectrograph optics, placed after the DLC in the optical train. Rather than proposing a monolithic spectrograph that would disperse light over the entire DICER focal plane, we instead imagined five separate SPICA-style spectrographs to tile the long, narrow, and curved focal plane as pictured. The five spectrograph design alleviates the need for unacceptably large immersion gratings and lenses. Light passes through the (curved) focal plane before entering the immersion grating wedge through its smallest rectangular side. It then reflects back out of the immersion grating wedge in the opposite direction and is picked off with a set of five mirrors in the focal plane arc before going through the corrective focusing optics that puts the spectrum onto five detectors (green squares in Figure~\ref{fig:dicer_inside_3}). 

\clearpage

The spectrum of an exoplanet will extend along the direction of the row of five detectors pictured in Figure~\ref{fig:dicer_inside_3}. The exoplanet spectrum will cover 2" of the sky in the vertical direction in the picture. With the $R \sim 100,000$ immersion spectrograph, the background zodiacal and exozodiacal light that is detected along with the light from the exoplanet comes from a portion of the sky that is four square arcseconds.

\subsection{Signal Modulation}

In DLC, the angular resolution enhancement of the POG is only realized in the direction of dispersion (because the active area of the grating is longer in that direction), so whether or not a close companion is resolved will depend upon the orientation of the grating relative to the angular separation of the host-star and close companion. If the dispersion direction of the grating is anti-aligned (90 degrees offset) from the angular separation of the host-star and close companion, the close companion will, in general, remain unresolved (provided that the angular separation is smaller than the diffraction limited resolution of the secondary telescopes). If the close companion is unresolved in a DLC configuration, the phase shifted close-companion PSFs overlap and are nulled along with the host-star light.

\begin{figure}
\begin{center}
\begin{tabular}{c}
\includegraphics[width=1\linewidth]{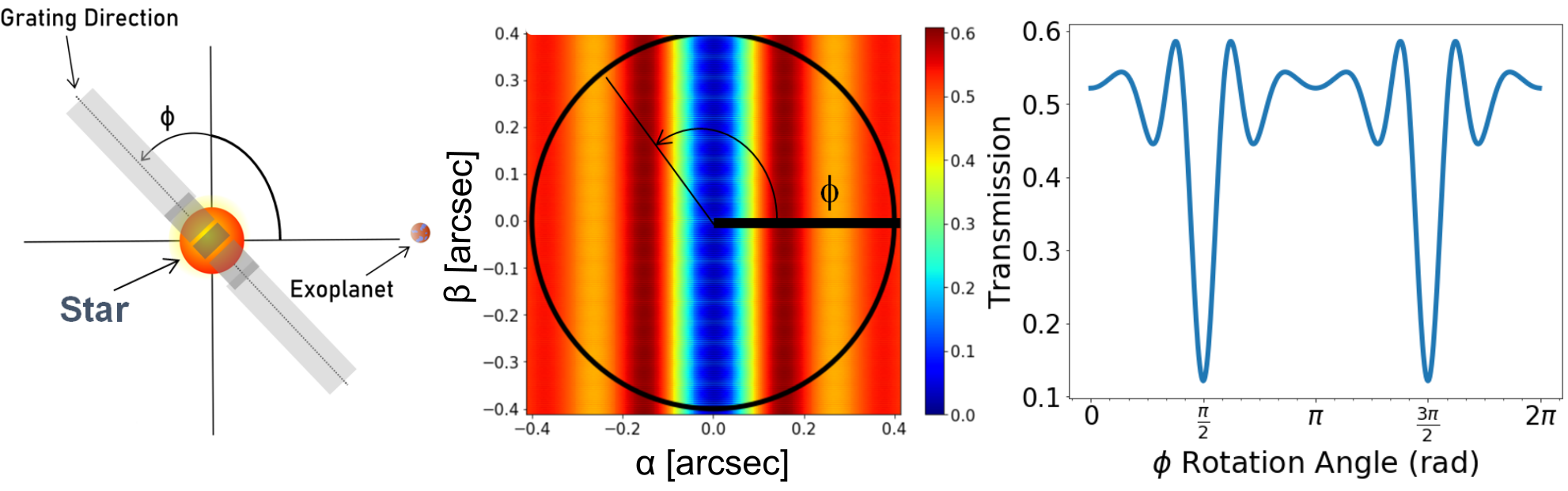}
\end{tabular}
\end{center}
\caption 
{ \label{fig:modu}
Illustration of the DICER observatory utilizing DLC for modulation of an exoplanet signal. If the entire satellite is rotated while still staring at the host-star (left), the exoplanet traces a circle in the 2-dimensional transmission map $\omega(\alpha,\beta)$ (center), resulting in a modulated exoplanet signal (right).
}
\end{figure} 

The one-dimensional resolution of DLC permits the use of signal modulation, an observational mode that is not realized with conventional AIC, and more closely resembles methods employed by other nulling interferometer architectures such as the original Bracewell configuration \cite{BRACEWELL1979136}. Using DICER as an example, if the entire observatory is slowly rotated, so the exoplanet position traces out a circle in the 2-dimensional transmission map $\omega(\alpha,\beta)$ shown in the center panel of Figure \ref{fig:modu}. The resulting exoplanet signal will then possess a characteristic modulation which depends on the angular separation of the exoplanet from the host star (radius of the circle traced in $\omega(\alpha,\beta)$).

Many applications seeking to characterize near-Earth exoplanets have proposed the use of signal modulation to significantly improve detection efficiency against noisy backgrounds. Past concepts such as Darwin\cite{LUND2001137} and the Terrestrial Planet Finder (TPF)\cite{TPF}, as well as the newly conceived Large Interferometer For Exoplanets (LIFE)\cite{refId0,refId2}, propose to achieve signal modulation by slow rotation of a space observatory. 

\subsection{Finding Exoplanets}

The development of an optimal survey strategy and detailed throughput information for each observation is beyond the scope of this work. Here, we present only a simple calculation to elucidate the capabilities of this system.

We selected from ExoCat-1 \cite{2015arXiv151001731T} a list of 15 main sequence stars within 8 pc of the Sun, that are either single systems or have an angular separation of 1" or more from any companion star, and for which the surface temperature is in the range $5100<T_{\it eff}<6600$. These stars include one F star, 9 G stars, and 5 early K stars. DICER is able to detect many planets around a wider range of host stars, but targeting more distant or cooler stars reduces the angular separation of the exoplanet and host star so that some of the inner planets for these hosts cannot be identified.

To discover exoplanets around these 15 nearby, Sun-like stars, DICER would stare directly at the host star (which is nulled with DLC) while rotating. The rotation is necessary because we do not a priori know the position angle of the planet with respect to the host star, and we only achieve 0.1" resolution between the exoplanet and host star in the long axis of the grating. Simply put, if a planet is between 0.1" and 1" from the host star, the exoplanet light will be resolved and detected when the long axis of the grating aligns with the star-planet position angle, and it will be nulled when the short axis of the grating aligns with the star-planet position angle. As the telescope rotates, the signal appears and disappears, allowing us to determine the current position angle of the exoplanet and an estimate of the angular separation from the host star. The technique of modulating the exoplanet signal through telescope rotation is employed for exoplanet discovery with DICER, and also with the notional Large Interferometer For Exoplanets \cite{2022A&A...664A..21Q,2022A&A...664A..22D}, which uses an array of smaller space telescopes that must fly in formation to very high precision. 

\begin{figure}
\begin{center}
\includegraphics[width=1\linewidth]{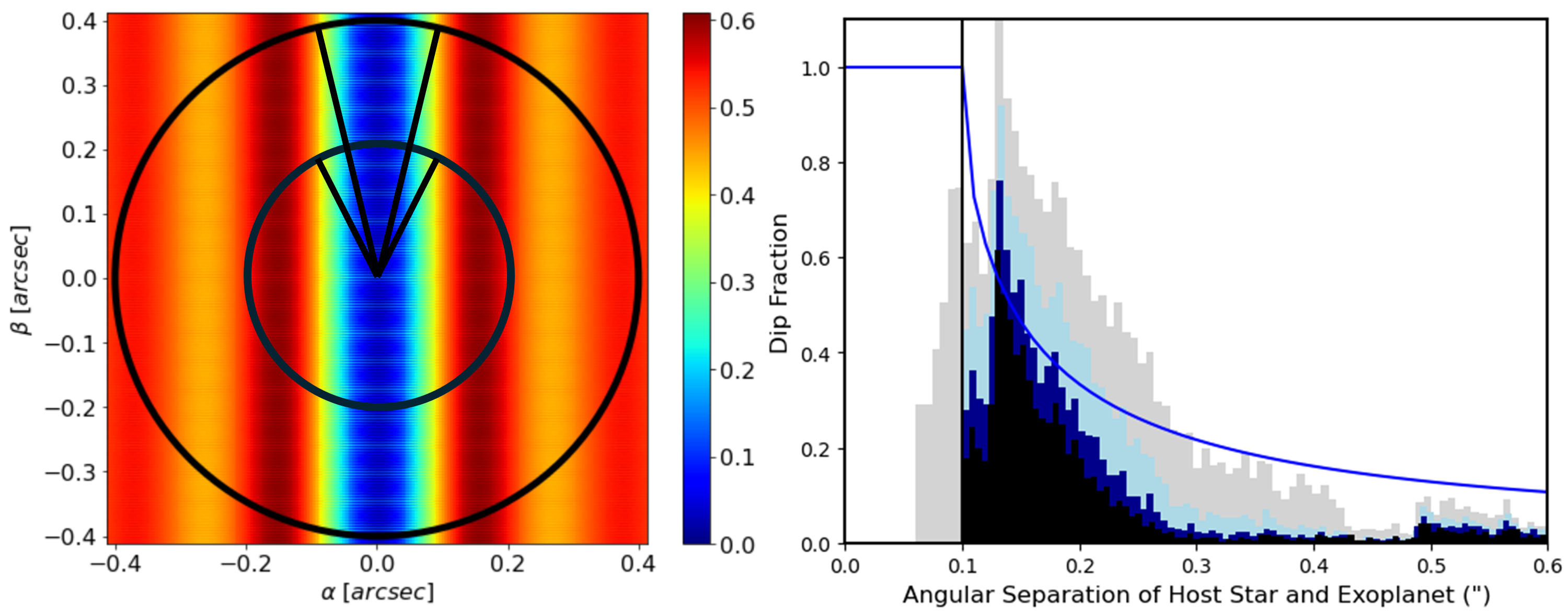}
\end{center}
\caption{The effect of transmission on planet detection. Color in the left panel shows the transmission fraction (a factor of 0.5 is included to account for light loss in the DLC) as a function of the relative sky position of the plant with respect to the star. As the telescope rotates while staring at the host star, the planet traverses the transmission plane in a circle with a radius given by the angular separation between the star and the planet. For smaller angular separations, the planet spends a larger fraction of the time in the low transmission (blue) state. The blue line in the right figure shows the fraction of the time the planet is in low transmission. To detect the planet, we must detect both the low transmission state and the high transmission state; for a given apparent brightness, an exoplanet is easiest to detect if the angular separation is 0.14", and is much more difficult to detect if it is closer than 0.1" to the host star because it stays in the low transmission state for the entire rotation. The grey histogram shows the distribution (scaled down by a factor of 500 to fit on the graph) of maximum separation between each star and simulated exoplanet that is reached during a full orbit, for the 13,624 simulated exoplanets. The light blue, dark blue, and black histograms show the maximum star-planet separation for the 1$\sigma$ detections (8348), the 3$\sigma$ detections (5539), and the 5$\sigma (4169)$ detections, respectively, for 80 days of observation per star.} \label{fig:dipfigure}
\end{figure}

The transmission function, and the path that the planet takes through the transmission function as DICER rotates, is shown in the left panel of Figure~\ref{fig:dipfigure}. Repeat observations, at times when the exoplanets are at different positions in their orbital paths, would allow us to trace planetary orbits and identify planets in edge-on or eccentric systems that are sometimes too close to the host star to detect. 

\begin{figure}[h!]
\begin{center}
\includegraphics[width=1\linewidth]{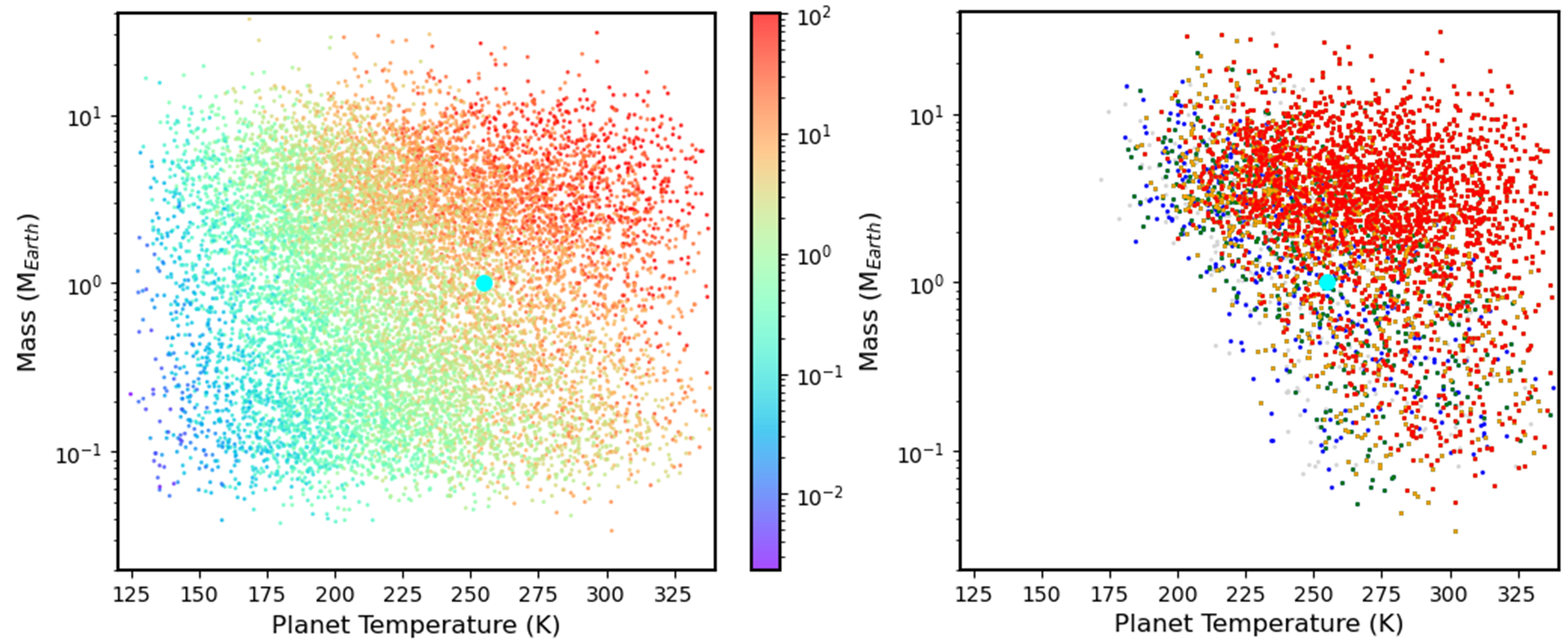}
\end{center}
\caption{Simulated habitable planets. We show in the left panel the masses and surface temperatures of 13,624 planets in 1000 simulations of habitable planets around 15 Sun-like stars within 8 pc. The colors of the points indicate the signal-to-noise ratio we would achieve in 80 days of observation of each host star with DICER. The right panel shows the 5$\sigma$ detections for 16 days (2424 planets, red), 32 days (3121 planets, orange), 48 days (3554 planets, green), 64 days (3883 planets, blue), and 80 days (4169 planets, light gray) of observation of each star. The blue cyan dot in each panel shows the mass and effective blackbody temperature of the Earth (255K, assuming and albedo of 0.3).}\label{fig:detections}
\end{figure}

Planets that are too close to the host star (hot) will not be resolved, and planets that are too far away (cold) will not emit significant light at 10$\mu$m; because of this, DICER preferentially finds only habitable zone planets.

We used the Bryson exoplanet population model\cite{2021AJ....161...36B}, as implemented in the P-pop exoplanet simulation tool \cite{2022A&A...668A..52K}, to create simulated habitable zone exoplanets around our 15 nearby, Sun-like stars. To achieve better statistics, the simulation was run 1000 times, simulating 1000 universes, so 13,624 habitable zone planets were generated (about one planet per star per universe). The masses and surface temperatures of the simulated planets are shown in the left panel of Figure~\ref{fig:detections}. The expectation from the model is that these 15 host stars host approximately 14 habitable planets.

We then calculated how many of the simulated exoplanets we would detect with DICER. Given the simulated exoplanet temperatures and radii, we calculated the distance to the host star, the blackbody flux at 10$\mu$m, the subtended solid angle of the exoplanet, and the planet flux as observed from the Earth. We then compared this signal with the background noise from stellar leakage and zodiacal light.

The primary background is zodiacal light in the Solar System or in the exoplanet's host solar system. The brightness of the local zodiacal cloud varies across the sky; it depends on the relative position of the Sun and Earth (and therefore depends on the time of year) and is typically brighter towards the Sun (where the dust is hotter) and near the plane of the ecliptic. Using IRAS data \cite{1998ApJ...500..525S} and the NASA Euclid background model tool \cite{zodiacallight}, we estimate that the local zodiacal light background, as observed from L2, varies between 10 and 20 MJy/sr. Unless the system is located very near the Celestial Equator, there is an optimal time of year when it can be observed with $\sim$10 MJy/sr of local zodiacal light background, which is what we assumed in the signal-to-noise calculation. 

Estimates of exozodiacal light background are very uncertain because zodiacal dust has not been observed in a representative set of systems. Dozens of systems have been found with zodiacal light that is orders of magnitude brighter than is found in the Solar System \cite{2007A&A...475..243D}, but these might be unusual systems; the exozodiacal light from most systems is too faint to detect. It is also unclear how much of the exozodiacal light will be nulled along with the host star. An important point about the exozodiacal light is that it will not be uniform, and therefore there is an extra importance above a straight signal-to-noise calculation to make observations with high resolution. While DLC nulls the central 1" by 0.1" region, the single wavelength planet PSFs are 2" by 0.2". However, because we cannot resolve wavelengths to the resolution of the POGs, light from the exoplanet is spread over a 2" by 2" region. In our signal-to-noise estimates, we assumed that the exozodiacal light in a 2" by 2" region around the exoplanet has twice the surface brightness as the minimum local zodiacal light. This is a bold estimate, but is justified by the following observations: the whole range of local zodiacal surface brightness is $10-20$ MJy/sr, including towards the Sun, towards the plane of the ecliptic, and looking out of the plane. Since the surface brightness of dust looking through a disk is twice as much as looking out of a disk from the midplane, and that is the same as the high numbers in the Solar System, and surface brightness is independent of distance, this estimate makes some sense. It is possible that the exozodiacal dust does not cover the whole $2" \times 2"$ background region for some systems, or that the dust in other systems could be much more or much less than in the Solar System, or that much of the exozodiacal dust (including much of the brighter, hotter dust near the star) could be nulled along with the host star light. These affects could make our assumption either optimistic or pessimistic, but we use 30 MJy/sr for the total zodiacal light background in the absence of real information about exozodiacal light.

The stellar leakage was calculated according to Equation~\ref{eqn:leaky}. We did not include star background from pointing jitter or residual optical path difference, because for our data these are much smaller than either the zodiacal light background or the stellar leakage; it is not expected that this oversight will significantly affect our results.

The optimal transmission function for the system was assumed to be 25\%, including 50\% from loss in the DLC beamsplitter and another 50\% due to other optical elements as was achieved for JWST \cite{2022SPIE12180E..0XG}. This might seem optimistic due to the large number of optical elements (for example the POG and immersion spectrograph), but the losses from these might be offset by the advantages of operating in an extremely narrow band along with improved detector efficiency from superconducting nanowire single-photon detectors \cite{2021JATIS...7a1004W, 2021APLP....6e6101V, 2022JLwT...40.7578L} and transition edge sensor bolometers \cite{2019APLP....4e6103H, 2021JATIS...7a1005N} which can detect single photons and exhibit virtually zero dark counts ($\sim$1 ct/day).

Figure~\ref{fig:dipfigure} shows the effect that the angular separation of the host star and exoplanet have on exoplanet detection. In the left panel, two circles corresponding to angular separations 0.2" and 0.4" are drawn. If the exoplanet orientation puts it within -0.1"$<\alpha<$0.1", then the exoplanet will be nulled. Exoplanets that are closer to the host star will be nulled for a larger fraction of the rotation. We have made the assumption that in order to detect the exoplanet we need to detect both the exoplanet (so it must be farther than 0.1" from the host star) {\it and} the dip in transmission as the exoplanet passes through $\alpha=0"$. We therefore calculated the signal-to-noise ratio (SNR) separately for the fraction of time the exoplanet would be detectable and the fraction of time the exoplanet would not be detectable; the right panel of Figure~\ref{fig:dipfigure} shows the fraction of a full rotation that an exoplanet is not detectable ($2\sin^{-1}(0.1"/\phi)/\pi$, where $\phi$ is the angle between the host star and exoplanet). We estimated the SNR for detecting an exoplanet as the minimum of the SNR for the detection part of the DICER rotation, and the calculated SNR for the period of time (exposure time multiplied by the fraction of time the exoplanet is undetectable) that the exoplanet has $|\alpha|<0.1"$.

Using this model, a 256K blackbody planet with the radius of the Earth, at a distance of 7.5pc, would produce $\sim$140 photons per hour. (Though note that the actual Earth emission is somewhat higher at 10$\mu$m than the blackbody emission for an an equivalent total emission object with a temperature of 256K.) The zodiacal light background is $\sim$1,200,000 photons per hour. And the stellar leakage for a G star at 7.5pc could produce 6000 photons per hour. Both the stellar leakage and exoplanet flux vary wildly based on distance as well as planet and star properties.

We estimated the output of a very simple survey strategy, in which each star is observed for the same amount of time. In practice, we expect the time to completion (or alternatively the number of identified exoplanets) could be significantly improved with an adaptive approach that uses a different exposure time based on the host star's distance and temperature, and that used the results of previous observations of each star to inform future observations.

We imagined exposing each host star for 16 days, while smoothly rotating DICER and staring at the star. Observing 15 stars requires 8 months of exposure time (but will take longer than this on the clock). We then observe each star five times over the course of 3.3 years; this both allows us to see planets that are sometimes more than and sometimes less than 0.1" from the host star and to track planets through their orbits. 

We imagine spending about three years of exposure time searching for habitable exoplanets. After the first 8 months, the planet model suggests we expect to find 2.4 habitable planets around Sun-like stars that are closer to the Sun than 8pc. After another 8 months of exposure time, we expect to find 3.1 habitable exoplanets. The planet model predicts that after a full 80 days of exposure time per star, we would find a approximately 4.2 exoplanets, or 30\% of the simulated exoplanets. DICER preferentially detects larger and warmer exoplanets in the habitable zone, as shown in Figure~\ref{fig:detections}. The angular separations of the 1$\sigma$, 3$\sigma$, and 5$\sigma$ detections for 80 days of observations per star are shown in the right panel of Figure~\ref{fig:dipfigure}. Note that there are some planets with maximum separations from their host stars of less that 0.1" that cannot be observed regardless of the amount of observing time.

\subsection{Measuring Biomarkers}
Once a habitable exoplanet is discovered, the next step is to search its spectrum for signs of life. One important indication of life is the presence of oxygen in the atmosphere. Although it is possible to make oxygen in planetary atmospheres without life, abiotic mechanisms are not expected to sustain oxygen in atmospheres of Earth-like habitable planets orbiting Sun-like stars for a sustained period of time \cite{2017AsBio..17.1022M}. Because ozone found in the Earth’s atmosphere is a result of photosynthesis, finding ozone in the atmosphere of an Earth-like planet would be a strong indicator of life and in particular of life that obtains energy from the host star through photosynthesis (on Earth photosynthesis occurs in vegetation and micro-organisms).

\begin{figure}[h!]
\begin{center}
\includegraphics[width=0.4\linewidth]{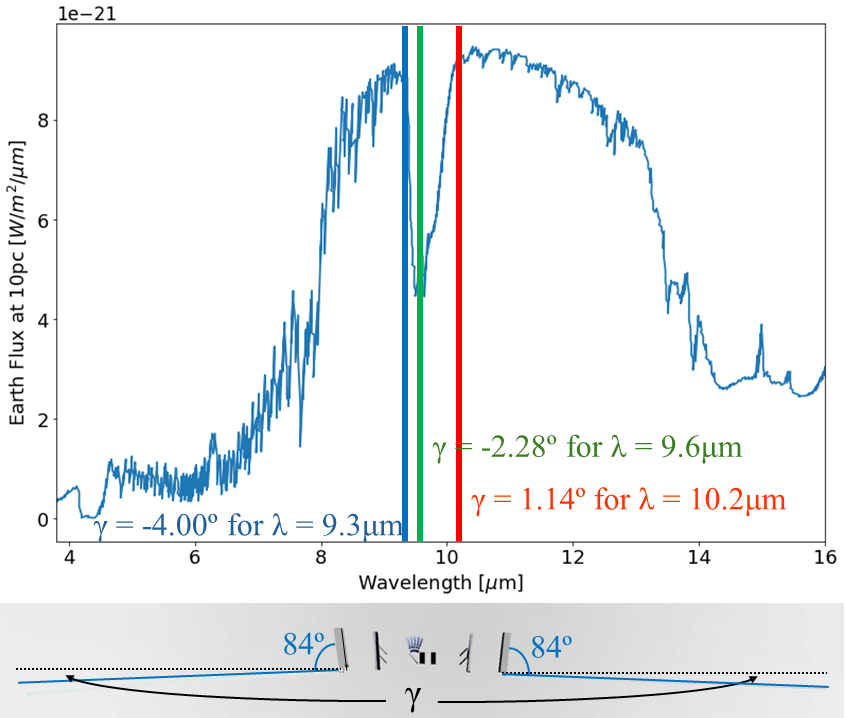}
\end{center}
\caption{Ozone detection by tilting the grating. The top panel shows the flux the Earth would produce if observed from a distance of 10 pc, using data from Robinson et al.\cite{2011ApJ...741...51R} Because the ozone absorption band at 9.6$\mu$m is much broader than the DICER bandpass, we observe the line by taking data at 10.2$\mu$m (the wavelength used for planet detection), 9.6$\mu$m (the wavelength at the minimum of the absorption band), and 9.3$\mu$m (a wavelength on the short side of the absorption band). If ozone is present, the flux at 9.6$\mu$m will be much lower than the other two. The detection wavelength of DICER is adjusted by tilting the gratings and associated focusing optics so that the host star no longer has an angle of incidence of $0^\circ$ (lower panel). If the gratings are tilted down (away from the target object), as shown in blue lines in the lower panel, then the angle of incidence is negative, while the angle of exodus remains $84^\circ$. Note that the dotted horizontal black lines are shown only to indicate where the gratings would be at zero tilt. The grating equation allows us to solve for the central wavelength that is detected, given the tilt. The grating angles ($\gamma$) required for 9.3$\mu$m, 9.6$\mu$m, and 10.2$\mu$m are -4.00$^\circ$, -2.28$^\circ$, and $1.14^\circ$, respectively.}\label{fig:DICERtilt}
\end{figure}

Our DICER design cannot measure the presence of the 9.6 $\mu$m ozone line in one observation, because the absorption line is about one micron wide (Figure~\ref{fig:DICERtilt}), and the DICER passband is only about 14.7 nm wide. However, we can tilt the gratings (and the attached secondary optics) so that the exoplanet’s angle of incidence to the grating changes; the exodus angle which is collected by the secondary telescopes remains the same. The detected wavelength therefore changes according to the grating equation. 

In order to sample the spectrum over wavelengths between 9.3 $\mu$m and 10.2 $\mu$m, the gratings need to tilt over a $5^\circ$ range (with the specified grating pitch, the required tilt is from $-4.0^\circ$ to $1.1^\circ$, as shown in Figure~\ref{fig:DICERtilt}). This will also require adjustments to the filter passbands, coronagraph, and input to the spectrograph that seem achievable but have not been modeled here.

To determine the time required to follow up 4 exoplanets, we calculated the time to reach a 10 sigma detection, under the assumption that we know where the planet is, so DICER doesn't need to rotate to find the signal. At 9.3$\mu$m we assume that the planet flux and backgrounds are the same as were used to calculate the SNR for detecting planets; for the average simulated planet, 25 days are required to obtain a 10$\sigma$ detection. At 9.6$\mu$m we assume that the planet flux is half as high because that is where the minimum of O$_3$ absorption line is, so on average 101 days are required. The total follow-up requires 126 days for each of 4 objects, which is 1.4 years of exposure time.

Combining 3.3 years of exposure time for finding exoplanets and 1.4 years of exposure time for spectroscopy, and assuming the telescope is exposing 70\% of the time, a 7-year mission would find $\sim 4$ exoplanets and determine whether they have ozone absorption in their spectrum.

\section{Discussion}

The DICER application, at least on paper, marginally works. Although our worked example did not find {\it all} of the habitable planets that were modeled around 15 Sun-like stars within 8 pc of the Sun, which was our original aim. However, it could plausibly find {\it some} nearby, habitable planets and search for ozone absorption in the spectrum. Because of the complexity of the system that would be required to achieve this, it is unlikely that this will turn out to be the best method for finding nearby exoplanets. 

For this reason, we have not attempted to address the myriad of engineering design considerations that are specific to the DICER application. For example, there are design considerations of how many segments the POGs should be composed of, and how those would be aligned. We have considered making each POG out of 1 meter square segments, because Xu et al. (2008)\cite{reactiveionetching} demonstrated a highly coherent diffracted wavefront from a meter-scale grating segment, thus demonstrating that a grating of this size is plausibly feasible. Presumably, each segment would need to be mounted on actuators in a similar manner to JWST, and would require additional lateral adjustment with neighboring grating segments.  If each segment could be manufactured to the required precision, alignment with neighboring segments (to within an integer number of line spacings) may then allow a 10m coherent wavefront. 

Another common concern is the question of thermal stability requirements, including for the specific use case of DICER, which in practice depend on many design decisions that have not been made. Because the baseline DICER design presented here uses reflection gratings, we could imagine using a beryllium substrate akin to that used for the JWST mirrors. The linear coefficient of thermal expansion of beryllium at the temperatures of the JWST primary segments (35-55K)\cite{2023PASP..135d8001R}, ranges from approximately 0.05-0.16 $\mu$m/(m K)\cite{Corruccini1961ThermalEO}, which is about 100 times lower than the value at 298K. The hottest (55K) segments of the JWST primary are those located closest to the sunshield, while the coldest (35K) segments are the most distant. The 20K temperature gradient of JWST would result in a length difference on the order of 2$\mu$m. Ideally, the mounting orientation of our grating segments being parallel to the sunshield surface would result in a smaller temperature gradient. Smaller temperature gradient, combined with significantly reduced coefficient of linear thermal expansion at low temperatures, push the thermal stability requirements plausibly into the realm of feasibility. However, this is undeniably a rough estimation.

\begin{figure}[h!]
\begin{center}
\includegraphics[width=0.9\linewidth]{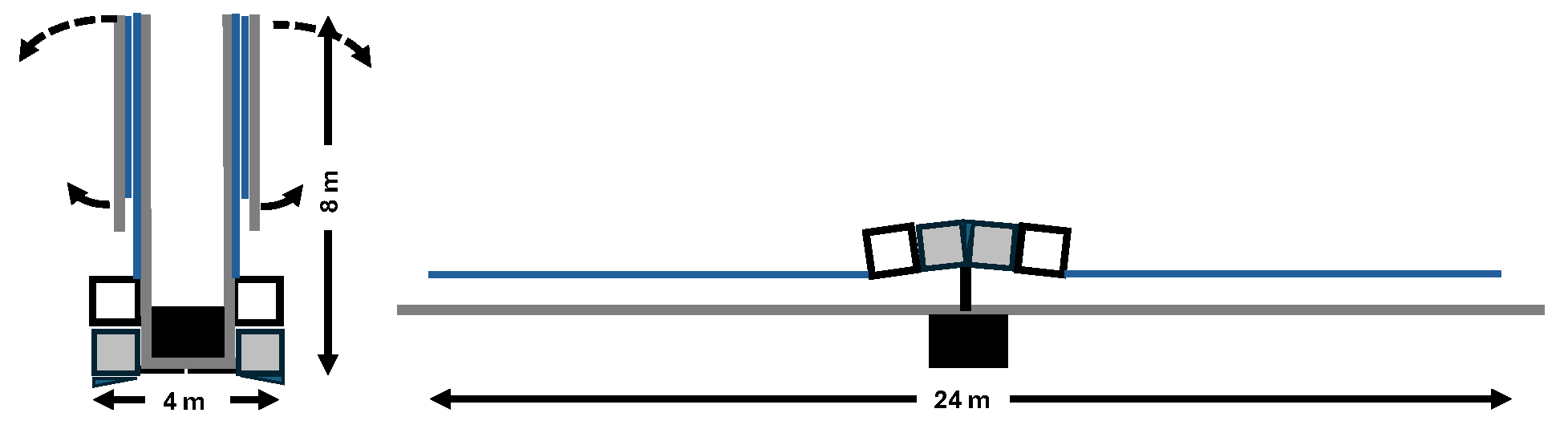}
\end{center}
\caption{Folding of DICER for launch. DICER was designed to fit inside a Falcon Heavy launch vehicle. The gratings (blue) and sunshield (thick gray lines) are initially folded (left diagram). These would unfold, and then the arms on the back of the spacecraft bus (black box) pivot so that the instrument packages (gray squares) come together. The secondary focusing optics (white squares) fold out with the instruments.}\label{fig:dicerfigure}
\end{figure}

There are a myriad of other concerns, including but not limited to: the feasibility of making ZnSe crystals that are 20cm in length, optimizing the throughput of DICER's secondary disperser, actively controlling the path length from both sides of the DLC, and fitting the gratings into a launch vehicle. Although we have not addressed all concerns, we have considered the DICER form factor during launch. DICER was sized to fit inside a Falcon Heavy launch vehicle, which has a payload volume that is 5.2m in diameter for 6.6m of height, and then the diameter gradually decreases to 1.3m over an additional 4.8m of height. Figure~\ref{fig:dicerfigure} shows diagrammatically how our design would fold up to a volume that is about 4m$\times$1m$\times$8.2m, which easily fits within the payload volume. The gratings (blue lines) that we had originally hoped could be launched as rigid structures had to be folded up to fit inside. The design could be scaled up about 20\% in grating length and secondary telescope size and still plausibly fit inside the launch vehicle in this configuration.

What is important about the DICER application is that it demonstrates the possibilities and limitations of a DLC design. Fundamentally, DLC produces extremely high resolution spectroscopy, combined with the ability to resolve and extinguish bright point sources. Although the spectroscopy is only obtained over a very small range, the wavelength range over which spectra are obtained can be tuned by varying the angles of the POGs. The system will in principle work for any wavelength range for which gratings can be manufactured and photons detected, including UV, optical, and infrared wavelength ranges.

Since binary stars are much brighter than any background zodiacal light, we could apply DLC to finding and characterizing faint companions to Galactic stars without the need for a secondary disperser. We could possibly study stellar mass loss on the giant branch and its relationship to binarity, and a myriad of other processes that effect stellar evolution. Similarly, applications to the study of protoplanetary disks and disks around active galactic nuclei should be considered.

Another possible application for DLC is in obtaining extremely high resolution spectra of subsets of stellar disks. Obtaining extremely accurate line-of-sight velocities for stars is important for exoplanet detection and characterization, and for making direct measurements of acceleration (allowing us to map the gravitational field, which in the Milky Way is primarily due to dark matter). These measurements are currently limited by convection and other surface properties of stars. With DLC technology, it might be possible to obtain extremely high resolution spectra of nearby stars, and probe the spatial distribution of radial velocity variation to study bulk flows on the surfaces of stars as a function of position on the surface.

\section{Conclusion}

We present a design for the Dispersion Leverage Coronagraph, a novel coronagraph architecture that is specially designed for use on POG telescopes, and explore its application to finding nearby, habitable planets with DICER. DLC utilizes the achromatic optical train of AIC to null a host star spectrum across the entire focal plane, though DLC will likely benefit from the use of periscope nulling techniques rather than the through-focus nulling utilized by AIC. Close companion (e.g. exoplanets or binary stars) spectra are preserved in DLC through spectral (rather than geometric) asymmetry in the focal plane. Here we have derived many important coronagraphic metrics for DLC:
\begin{itemize}
    \item {The transmission map of DLC exhibits the angular resolution of an aperture of width 2D, and therefore a diffraction limit of $2L/\lambda$, where $L$ is the grating length (since PSF positions in the focal plane are determined by the grating resolution\cite{tel}). This is similar to results found for AIC (and, in fact, the resolution in the oblique $\beta$ direction is identical to AIC, with a resolution of $2D/\lambda$, where $D$ refers to the diameter of the telescope and also the width of the POG), and highlights an advantage of this coronagraph design. Thus, to resolve a planet and star 10 parsecs away, our double grating design of length 10 meters is sufficient, whereas normally a much heavier 20 meter diameter monolithic mirror would be required.}
\item {Under ideal circumstances with a star at perfectly normal incidence to the diffraction grating ($\alpha=0$), the DLC can effectively null the image across the entire focal plane by aligning the stellar PSF at every wavelength position. This is in contrast to other coronagraphs like AIC which only null over the middle of the focal plane. Although DLC is similar to AIC in that it nulls only at the center of the FOV, the nulling is asymmetric, and will partially null sufficiently close companions when the grating is anti-aligned with the angular separation of the companion and host-star.}
\item{Overlaying the finite angular extent of the stellar disk on the transmission function allowed for derivation of stellar leakage components $\omega$:
\begin{equation}
\omega(r_*) \approx I_0 r_*^2  \frac{D^4\pi^2}{6\lambda^2} \left( 1 + (1 - \lambda^2/p^2)^{-1} - r_*^2 \: \left( \frac{D\pi}{3\lambda} \right)^2 (1-\lambda^2/p^2)^{-1} \right).
\end{equation}
 For the numerical example of the current DICER benchmark values (10m gratings with 1m secondary telescopes placed at an exodus angle of $84^\circ$, observing a Sun-like star at a distance of 10pc), stellar leakage was shown to impose a $10^5$ rejection limit.}
\item{The effects of pointing jitter of the telescope on transmission is more dependent on the dispersion angle ($\alpha$) than the oblique angle ($\beta$) according to the relation:
\begin{equation}
\left< Rej \right> = \frac{1}{\left< \alpha^2 \right>\frac{2\pi^2 D^2p^2}{3\lambda^2p^2-3\lambda^4} + \left< \beta^2 \right> \frac{2\pi^2 D^2}{3\lambda^2}}.
\end{equation}
 For any POG telescope, there must be tighter requirements on alpha than beta to achieve the proper stellar flux rejection level. Again using the DICER values for a numerical example, fine guidance with similar accuracy to JWST results in a factor of $10^4$ starlight suppression limited by pointing jitter.}
\item{DLC enables higher resolution in one dimension due to the use of POGs for optical leverage. This one-dimensional resolution leaves open the possibility for signal modulation techniques frequently utilized by other coronagraph architectures.}
\item{When designing a DLC arrangement for close-companion spectroscopy with a set angular resolution limit, the bandwidth of the optical system is invariant with respect to the grating length/exodus angle, and is dependent only upon the FOV of the secondary telescope.}
\end{itemize}

The results presented here are the first step to understanding the full capabilities of DLC. We have shown that for the notional DICER architecture, DLC would effectively suppress starlight to levels potentially suitable (similar to the Earth/Sun contrast ratio) for exoplanet detection near 10$\mu$m wavelength. 

Using the benchmark DICER design, we plausibly expect to find $\sim 4$ habitable exoplanets around Sun-like stars that are closer than 8pc, and determine whether they have ozone absorption in their spectra, within a seven year mission. While the DICER design makes it possible to launch a 20m-class infrared telescope in a Falcon Heavy launch vehicle, there are many other engineering hurdles that must be overcome to make this concept a reality.

Most notably, DICER is much more complex than DLC alone because it requires a very ambitious second disperser to separate exoplanet light from the much brighter Zodiacal light background; and even with that second disperser the exoplanet is not separated from the background with the full resolution of the 10 meter grating. (Note that the angular region that is extinguished in the DLC does benefit from the full diffraction limit of the gratings, with or without a second disperser.) This suggests that DLC might be more successful in applications with brighter objects with lower backgrounds. The extremely high native spectral resolution is also a unique feature to be exploited. 

In principle, the DLC may be useful for any application requiring extremely high resolution, close-companion spectroscopy. It can be applied in the UV, visible or infrared wavelength ranges. Use cases currently under consideration are exoplanet discovery by the radial velocity method, disks around active galactic nuclei, binary stars, protoplanetary disks, and high resolution spectroscopy for precise measurement of acceleration. Because the unusual capabilities of DLC have never before been available, the best use cases may not have been previously suggested because they were thought to be impossible.

\appendix
\section{\label{sect:ident} Sinc Integral}

Required in the derivation of the integral solution in Eqn. \ref{eqn:wholef} is the indefinite integral:

\begin{equation*}
\int \textrm{sinc}(bx+c) \textrm{sinc}(bx+a)dx = \frac{1}{2\pi^2b(a-c)}\Big\{  \textrm{cos}(\pi(a-c))\Big(\textrm{Ci}(2\pi|a+bx|) - \textrm{Ci}(2\pi|c+bx|)\Big) +
\end{equation*}

\begin{equation}
\label{eqn:intensity}
    \textrm{sin}(\pi(a-c))\Big(\textrm{Si}(2\pi(a+bx)) + \textrm{Si}(2\pi(c+bx))\Big)  -  \textrm{cos}(\pi(a-c))\Big(\textrm{log}|a+bx| - \textrm{log}|c+bx|\Big)  \Big\} + \textrm{const.}
\end{equation}

Here, $\textrm{Si}(x) = \int_{0}^{x} \textrm{sin}(x')/x' dx'$ and $\textrm{Ci}(x) = -\int_{x}^{\infty} \textrm{cos}(x')/x' dx'$


\subsection* {Data Availability}
The models used to generate Figures \ref{fig:detections} \& \ref{fig:dipfigure} can be accessed using the code resources found in the Bryson\cite{2021AJ....161...36B} and LIFE\cite{2022A&A...664A..21Q,2022A&A...664A..22D}  references. Any reader seeking our particular dataset, obtained by running these models using custom parameters, may contact the first or second author for access.

\subsection* {Acknowledgments}
This paper was supported by Phase I grant 80NSSC23K0588 from NASA Innovative Advanced Concepts (NIAC), and Manit Limlamai. Leaf Swordy was supported by a fellowship from the NASA/NY Space Grant.

\subsection* {Disclosures}
The authors declare there are no financial interests, commercial affiliations, or other potential conflicts of interest that have influenced the objectivity of this research or the writing of this paper.





\vspace{2ex}\noindent\textbf{Leaf Swordy}  is a postdoctoral researcher at the National Institute of Standards and Technology (NIST), working in the field of superconducting detector R\&D. He received his BS, MS, and PhD in physics from Rensselaer Polytechnic Institute in 2018, 2019, and 2024 respectively. His background includes high-purity liquid noble gas systems and general R\&D for the XENON1T and nEXO experiments, and the optical design of theoretical telescope architectures utilizing holographic primaries; the most notable example being the Diffractive Interfero Coronagraph Exoplanet Resolver.

\vspace{2ex}\noindent\textbf{Prof. Heidi Jo Newberg}
 is a Professor in the Department of Physics, Applied
Physics, and Astronomy at Rensselaer Polytechnic Institute. She made
significant contributions to the Supernova Cosmology Project, the Sloan
Digital Sky Survey, the Chinese LAMOST project, and currently runs the
MilkyWay@home volunteer supercomputer. She is a Fellow of the American
Physical Society and has published in diverse areas of astrophysics
including discovery of tidal streams and substructure in the Milky Way
halo, discovery of disequilibrium in the Milky Way disk, properties of
stars, astronomical surveys, supernova searches, discovery of dark
energy, the constraining the spatial distribution of dark matter, and now Dittoscopes.

\vspace{2ex}\noindent\textbf{Becket Hill} is a physics doctoral student at the University of Illinois Urbana-Champaign. As an undergraduate, he made calculations for transmission and signal modulation of the DICER telescope, as well as contributing to research on the reactor antineutrino anomaly at Brookhaven National Lab. He is currently studying the magnetic properties of ferromagnets in order to search for exotic spin-dependent forces.

\vspace{2ex}\noindent\textbf{Dr. Richard K. Barry} is a research astronomer for NASA in the Laboratory for Exoplanets and Stellar Astrophysics.  With a background and degrees in physics, instrument design and electromechanical engineering, Dr. Barry has contributed to many high-profile space missions including Hubble Space Telescope, X-ray Timing Explorer, Transition Region and Coronal Explorer, James Webb Space Telescope and Roman Space Telescope.

\vspace{2ex}\noindent\textbf{Marina Cousins} is an undergraduate at Rensselaer Polytechnic Institute, majoring in mechanical engineering and minoring in astrophysics. She contributed to the science plan and thermal analysis of the Diffractive Interfero Coronagraph Exoplanet Resolver telescope concept. She is currently an intern at NASA Marshall Space Flight Center’s Optical Systems Design and Fabrication branch supporting development of novel design methods to reduce weight in structures that are driven by vibrations.

\vspace{2ex}\noindent\textbf{Kerrigan Nish} is an undergraduate physics major at Rensselaer Polytechnic Institute. She contributed to simulations of exoplanet detection using the Diffractive Interfero Chronograph Exoplanet Resolver, and is currently exploring a range of physics research experiences.

\vspace{2ex}\noindent\textbf{Frank Ravizza} is an Optical Engineer with 25 years of experience at Lawrence Livermore National Laboratory (LLNL). He received a B.S. and M.S. in Optical Science and Engineering from UC Davis and the University of Arizona (respectively). Frank contributed advances in optical metrology critical to the sustained operation of the National Ignition Facility culminating in fusion energy gain from a laser driven inertial confinement target in 2022. Additionally, Frank has multiphysics modeling expertise of intracavity aberrations in unstable laser resonators for high average power applications. Currently, Frank is an R\&D leader within LLNL’s space program, where he is advancing large aperture monolithic telescopes to enable lower cost space missions for astronomers and astrophysicists. 

\vspace{2ex}\noindent\textbf{Sarah Rickborn} is an undergraduate aerospace engineering and physics dual major at Rensselaer Polytechnic Institute. Her key focus is in the aerospace and astrophysics fields. She has worked on the science plan for the Diffractive Interfero Coronagraph Exoplanet Resolver for the past two years.

\end{document}